\documentclass[aps,onecolumn,showpacs,groupedaddress]{revtex4}
\usepackage{epsfig}
\usepackage{bm}
\usepackage{graphics}
\usepackage{bm}
\usepackage{amsmath,amssymb}

\newcommand{\etal}{{\it et al.}}

\def\shiftleft#1{#1\llap{#1\hskip 0.04em}}
\def\shiftdown#1{#1\llap{\lower.04ex\hbox{#1}}}
\def\thick#1{\shiftdown{\shiftleft{#1}}}
\def\b#1{\thick{\hbox{$#1$}}}
\begin{document}
\preprint{02-02}
\title{A dressed-bag model study of the final-state $N\Delta$ interaction
        in pion-photoproduction processes off the deuteron}
\author{I.T. Obukhovsky}
\author{V.I. Kukulin}
\affiliation{Institute of Nuclear Physics, Moscow State University, 119899
                Moscow, Russia}
\author{Murat M. Kaskulov}
 \email{kaskulov@pit.physik.uni-tuebingen.de}
\author{Peter Grabmayr}
 \email{grabmayr@uni-tuebingen.de}
\affiliation{Physikalisches Institut, Universit\"at  T\"ubingen,
         D-72076 T\"ubingen, Germany}
\author{Amand Faessler}
\affiliation{Institut f\"ur Theoretische Physik, Universit\"at  T\"ubingen,
         D-72076 T\"ubingen, Germany}
\date{\today}

\begin{abstract}
The impact of the short-range $N\Delta$ interaction on the pion
 photoproduction processes off the deuteron in the $\Delta$-resonance region
 is studied in the framework of recently proposed dressed-bag model. A common
 dressing procedure for bare three- and six-quark states is used to describe
 both the pion decay widths of baryon resonances and the effective $NN$ (or
 $N\Delta$) interaction at short ranges related to the inner dressed-bag
 states. It is shown that the effect of short-range $N\Delta$ interaction for
 the forward-angle $\pi^\mathrm{o}$ photoproduction off the deuteron cannot be
 neglected.  The prospects for further development of the model to describe
 the short-range $NN$ (or $N\Delta$) correlations in the lightest nuclei are
 discussed.
\end{abstract}
\maketitle
\section{Introduction}
\label{sec:intro}

The $\Delta$-isobar degrees of freedom play a crucial role in the description
 of numerous processes in nuclear and hadron physics. In such a description
 the $\Delta$-isobar is either generated in the process itself or can be
 considered as ``preexisting'' in the wave function of the initial system. In
 any case we need to understand the dynamics for the $\Delta$-isobar in a
 nuclear system, which in turn requires the construction of an adequate
 model for the $N\Delta$ interaction. The long-range part of the $N\Delta$
 interaction is governed by the $\pi$-exchange, but its intermediate- and
 short-range parts are poorly known up to date, and thus it is an important
 topic for further studies. On the other hand, at low energies the
 $\Delta$-isobar due to its rather high mass should be taken in the far
 off-shell region, and thus, the short-range part of the $N\Delta$ interaction
 will be essential for the understanding of many low- and intermediate-energy
 processes.

Unfortunately, even for the $NN$ system which has been studied in much more
 details than the $N\Delta$ system the short-range part of the interaction is
 known rather poorly.  In the current $NN$ models the short-range part of the
 interaction (e.g. the high-momentum behaviour of vertex form factors in the
 meson-exchange models) is only fitted to the experimental $NN$ phase
 shifts. Because such a fit would be hardly possible for the $N\Delta$ system,
 one should resort to a consistent microscopic treatment of the short-range
 interaction within a realistic model with a common set of parameters for the
 $NN$ and $N\Delta$ systems. In our opinion any consistent microscopic
 treatment of short-range $NN$ and $N\Delta$ interactions should include both
 quark and meson degrees of freedom.

The quark approach to the short-range $NN$ interaction has a more than
 two-decade history during which many methods have been developed, e.g. the
 adiabatic~\cite{lib,obu,har,cve} and resonating group methods
 (RGM)~\cite{oka,fae}, quark compound bag (QCB) model~\cite{sim,bak}, hybrid
 models~\cite{kis}, etc. These models are in agreement with the $NN$ data at
 intermediate energies, but all of them treat the inner $NN$-system on a
 rather phenomenological level. Important meson field degrees of freedom remain
 absent.

In the recently proposed Moscow-T\"ubingen dressed-bag model~\cite{kuk} we
 start from the quark-configuration analysis of the RGM wave function
 including the configuration mixing at short range~\cite{kus} and take into
 consideration the proper meson cloud including 2$\pi$ and $\sigma$
 contributions of the bag-like six-quark component in the overlap
 region around 0.6~fm. Contrary to other models which exploit the
 meson cloud coupled to the quark core of the nucleon (e.g. the cloudy bag
 model~\cite{thom}) we have formulated our approach to the six-quark
 intermediate states using the Fock-column representation for the 
 vectors of state and we have proposed a new mechanism for the dressing of the
 six-quark bag with the $\sigma$-meson field. With three phenomenological
 parameters in each partial wave~$L$ the model describes the $NN$ scattering
 data for $L\leqslant 2$ up to 1~GeV with high precision including the tensor
 mixing and all the properties of the deuteron~\cite{kuk}. We assume that the
 model describes the short-range $N\Delta$ interaction as well. In both
 channels, $NN$ and $N\Delta$, the short-range interaction is determined by
 the same constraint originating from the Pauli exclusion principle.

\begin{figure}[hp]\centering
\epsfig{file=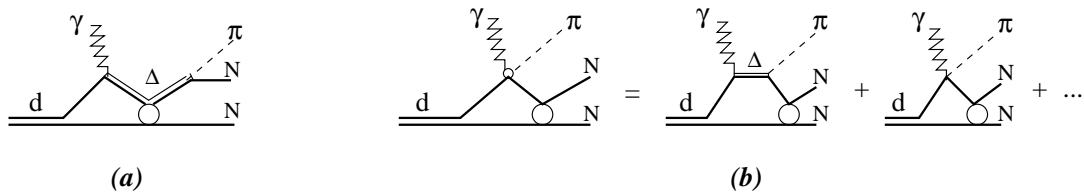,width=0.8\textwidth}
\caption{\label{f1}
 Loop diagrams of FSI for the ``direct'' short-range $N\!-\!\Delta$ 
interaction before the pion emission (a) and for the $NN$ rescattering 
after the pion emission (b)}
\end{figure}

Another key problem in this field is the possibility of tests of this part of
 the $N\Delta$ interaction. If one samples the $N\Delta$ wave function by
 quasielastic $(e,e^{\prime})$ knock-out of the $\Delta$-isobar (or of the
 nucleon) from the deuteron, it would be rather a study of a ``preexisting''
 $N\Delta$ component. Another method consists in the $\Delta$-isobar
 excitation by real photons and in a study of the off-shell $N\Delta$
 rescattering amplitude in the final state (diagram in Fig.~\ref{f1}(a)). We
 have selected the pion photoproduction reactions $d(\gamma,\pi^\mathrm{o}pn)$
 and $d(\gamma,\pi^-pp)$ in the $\Delta$-resonance region in a kinematic
 setting which maximises the pion kinetic energy. The corresponding
 forward-angle photoproduction cross sections are sensitive to the short-range
 part of the $N\Delta$ interaction.  To prove this statement we evaluate the
 contribution of the short-range $N\Delta$ final-state interaction (FSI) to
 the pion photoproduction cross section starting from the constituent quark
 model (CQM) for baryons and from the dressed-bag model~\cite{kuk} for $NN$
 and $N\Delta$ interactions.

The CQM describes with reasonable accuracy all the electro-magnetic and strong
 production processes at least for the free nucleon~\cite{gia,yao,ris}.  The
 results for free nucleons can readily be generalized for any baryon and for
 the six-quark component of any nucleon-baryon system.  Here we consider $\pi$
 photoproduction in the $\Delta$-resonance region, but such a consideration
 could be easily extended to $\eta$ photoproduction in the $S_{11}$-resonance
 region, or to other similar cases. Even though the CQM approach is limited in
 use by a non-relativistic harmonic-oscillator (h.o.) basis and a naive
 $^3P_0$ model of meson production, there is the advantage, that it allows us
 to embody all the relevant baryon resonances in the reaction mechanism with
 predictable amplitudes. The corresponding experimental data can be directly
 related to the internal structure of these resonances.

Up to now the role of the $N\Delta$ final state interaction (FSI) in the pion
 photoproduction off nuclei has been poorly investigated.  We recall that the
 $\Delta$-resonance region at $E_{\gamma}\approx\,$250-350~MeV is most
 convenient for such a study since the influence of other resonances,
 e.g. $P_{11}$(1440) or $S_{11}$(1535), is negligible in this region, and thus
 the main mechanism of photoproduction is a quasifree process with a virtual
 $\Delta$-isobar excitation.  It is well known that the contribution of this
 mechanism to the $\pi^\mathrm{o}$ production off the deuteron overestimates
 the forward-angle cross section~\cite{lag,are,lev1} by a factor of 1.5 to 2
 and that the FSI should play a decisive role in this
 region~\cite{kol}. Recent evaluations~\cite{lev2} have shown that taking into
 account the $pn$ FSI can bring theoretical predictions into better agreement
 with the data~\cite{kru}.

It is quite plausible that the $N\Delta$ FSI plays also a certain role in the
 $\pi^\mathrm{o}$ photoproduction along with the $NN$ FSI~\cite{lev2}, and if
 so, there will be a possibility to study the $N\Delta$ interaction in the
 photoproduction processes on nuclei. The contribution of the direct $N\Delta$
 interaction to pion production depends essentially on the off-shell behaviour
 of the $N\Delta$ amplitude because of the integration over the inner momenta
 in the loop diagram shown in Fig.~\ref{f1}(a).

Earlier attempts to take the direct $N\Delta$ interaction into account in
 terms of an $s$-channel resonance pole mechanism~\cite{lag} did not bring
 a detectable improvement in the description of the data. On the other hand
 there is a successful coupled-channel ($NN+N\Delta$) approach to pion
 production in NN collisions~\cite{sau}, and for the full study of the FSI
 problem it would be desirable to include a couple-channel description of both
 $NN$ and $N\Delta$ final states. Then the long-distance $N\Delta$
 correlations would be fully taken into consideration.

In a more simple approach. which is used here, the final-state $NN$
 interaction includes diagrams with $N\Delta$ intermediate states, as shown in
 Fig.~\ref{f1}(b) by the diagrammatic representation of this interaction.  The
 virtual $N\Delta$ channel is partially taken into account in the $NN$ FSI.
 Simplifying the problem in such a manner, we imply further that the diagram
 in Fig.~\ref{f1}(a) is related only to the ``direct'' short-range $N\Delta$
 interaction.

By the following prescriptions of the dressed-bag model we evaluate here the
 contribution of this ``direct'' short-range $N\Delta$ FSI to the differential
 cross section and show that this contribution is not small and cannot be
 neglected. As a first step of the study, it is demonstrated by comparison of
 the results of two simplified models for pion photoproduction off the
 deuteron: (i) the quasifree mechanism, (ii) the quasifree mechanism 
 modified by the
 $N\Delta$ FSI Born term.  A more realistic variation of the model considering
 both $NN$ and $N\Delta$ half-off-shell t-matrices in the FSI loop integrals
 is in progress.

In Sect.~\ref{sec:srinteraction} of this paper, the quark-configuration
 analysis of $NN$ and $N\Delta$ systems in the overlap region is briefly
 reviewed, and the constraints following from the Pauli exclusion principle
 are formulated. The effective short-range $N\Delta$ interaction is deduced
 from the dressed-bag model with the same method as for the $NN$ interaction
 in Ref.~\cite{kuk}. In Sect.~\ref{sec:nucleon}, pion photoproduction off the
 nucleon in the $\Delta$-resonance region is described in terms of the
 CQM. The dressing of a bare (three-quark) $\Delta$-isobar and other baryons
 is carried out with the $^3P_0$ model. On this basis the ``off-shell''
 $\Delta$ propagator is obtained.  In Sect.~\ref{sec:deuteron}, the results on
 pion photoproduction off the deuteron are discussed and the concluding
 remarks are given in Sect.~\ref{sec:conclusion}.


\section{Short-range $N\Delta$ interaction}
\label{sec:srinteraction}

Within the quark-model the short-range $N\Delta$ and $NN$ interactions are
 very similar to each other. In fact, in the lowest partial waves, in the even
 ($L=0,\,\,2$) and in the odd ($L=1,\,\,3$) states, quark configurations of
 the $NN$ and $N\Delta$ overlap have the same structure determined by the
 constraint following from the Pauli exclusion principle.  We consider here
 only the lowest even partial wave $L=0$ for the spin-isospin state $ST=10$ in
 the $NN$ channel and for $ST=11$ and $ST=21$ in the $N\Delta$ channel.  These
 quantum numbers are important for the pion photoproduction off the deuteron
 in the $\Delta$-resonance region.

\subsection{Configuration mixing in $NN$ and $N\Delta$ overlap regions}
\label{ssec: confmix}

\begin{figure}[hp]
\begin{center}
\epsfig{file=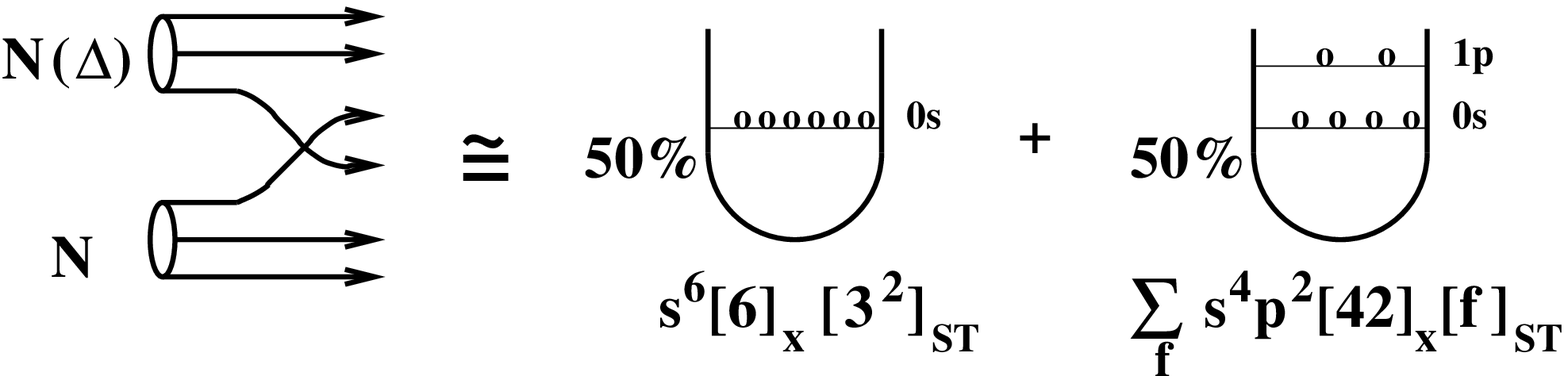,width=0.5\textwidth,clip}
\caption{\label{f2}
Configuration mixing at short range  in the
 $NN(ST{=}10)$ (or $N\Delta(ST{=}21)$) channel.}
\end{center}
\end{figure}

Microscopic evaluations~\cite{kus} made in the RGM approach lead to the
 following picture of configuration mixing in the overlap region as depicted
 in Fig.~\ref{f2}.
In both $NN(ST{=}10)$ and $N\Delta(ST{=}21)$ channels the overlap can be
 described by 
 a fully symmetric configuration $s^6\,[6]_{\scriptscriptstyle X}$ which has a 
 probability of $\approx$50\% in the overlap region $r\lesssim$0.6-1 fm;  
 the other 50\% being a superposition of states
 in 2$\hbar\omega$-excited configurations $s^4p^2$, which are characterized
 by a non-trivial coordinate Young tableaux $[42]_{\scriptscriptstyle X}$.
 However, in the other $N\Delta$ channel ($ST{=}11$) the fully symmetric state
 $s^6$ is  forbidden by the Pauli exclusion principle.

The states in 6q-configurations $s^6$ and $s^4p^2$ differ in their inner
structure:
\begin{itemize}
\item[(i)] The fully symmetric configuration $s^6$ is a bag-like state with
           its charge-spin structure strictly constrained by the Pauli
           exclusion principle. In both channels, $NN(ST{=}10)$ and
           $N\Delta(ST{=}21)$, the Young tableaux in the product of spin and
           isospin spaces $[3^2]_{\scriptscriptstyle ST}$ is fixed by the
           condition to be conjugate to the colour Young tableau
           $[2^3]_{\scriptscriptstyle C}$. The fully antisymmetric state
           $[1^6]_{\scriptscriptstyle CST}$ can only be constructed as an
           inner product of mutually conjugate states
           $[2^3]_{\scriptscriptstyle C}\circ [3^2]_{\scriptscriptstyle ST}$.
\item[(ii)] In the excited six-quark shell-model states  $s^4p^2$ of
           $[42]_{\scriptscriptstyle X}$ and  
           $[6]_{\scriptscriptstyle X}$ symmetry, the non-trivial
           coordinate Young tableau $[42]_{\scriptscriptstyle X}$ allows all
           the possible tableaux $[f_{\scriptscriptstyle ST}]$ in the inner
           product of spin and isospin spaces for both $NN(ST{=}10)$ and
           $N\Delta(ST{=}21)$ channels:
\begin{eqnarray}
NN(10):\qquad [f^{\scriptscriptstyle (N)}_{\scriptscriptstyle ST}]&=&
[42]_{\scriptscriptstyle S}\circ[3^2]_{\scriptscriptstyle T}=
[51]_{\scriptscriptstyle ST}+[3^2]_{\scriptscriptstyle ST}+
[41^2]_{\scriptscriptstyle ST}+[321]_{\scriptscriptstyle ST}+
[2^21^2]_{\scriptscriptstyle ST},\nonumber\\
N\Delta(21):\qquad
[f^{\scriptscriptstyle (\Delta)}_{\scriptscriptstyle ST}]&=&
[51]_{\scriptscriptstyle S}\circ[42]_{\scriptscriptstyle T}=
[51]_{\scriptscriptstyle ST}+[42]_{\scriptscriptstyle ST}+
[3^2]_{\scriptscriptstyle ST}+[41^2]_{\scriptscriptstyle ST}+
[321]_{\scriptscriptstyle ST}.\quad
\label{you}
\end{eqnarray}
           (This is also true for the other $N\Delta$ channel $ST{=}11$ which
           contains the product
           $[42]_{\scriptscriptstyle S}\circ[42]_{\scriptscriptstyle T}$).
\end{itemize}

Consequently, in the configuration $s^4p^2$ in each $ST$ channel there is a
 six-dimensional basis with vectors labelled by a set of Young tableaux
 $f=\{[f_{\scriptscriptstyle X}],\,[f_{\scriptscriptstyle ST}]\}$, where
 $[f_{\scriptscriptstyle X}]=[6],\,[42]$ and $[f_{\scriptscriptstyle ST}]$
 takes all the values $[f^{\scriptscriptstyle (N)}_{\scriptscriptstyle ST}]$
 or $[f^{\scriptscriptstyle (\Delta)}_{\scriptscriptstyle ST}]$ from
 Eq.(\ref{you}):
\begin{eqnarray}
d_f(s^4p^2)&=&|s^4p^2\,[f_{\scriptscriptstyle X}]\,
[f^{\scriptscriptstyle (N)}_{\scriptscriptstyle ST}](ST{=}10),L=0\,
\rangle\ \ \mathrm{and}\nonumber\\
D_f(s^4p^2)&=&|s^4p^2\,[f_{\scriptscriptstyle X}]\,
[f^{\scriptscriptstyle (\Delta)}_{\scriptscriptstyle ST}](ST{=}21),L=0\,
\rangle,
\label{df}
\end{eqnarray}
where $d_f$ and $D_f$ are the excited quark-shell-model configurations
with the $NN$ and $N\Delta$ quantum numbers respectively. 
This is a sufficiently large basis to
represent the main features of the $NN$ or $N\Delta$ channels at short
ranges. In Ref.~\cite{kus} it was also shown that some linear combinations of
the basis states (Eq.\ref{df}) are genuine $NN$ (or $N\Delta$) cluster-like
states, which can be referred to as the short-range wave functions
$\{\Psi_{\scriptscriptstyle NN}\}_{sr}$ and $\{\Psi_{\scriptscriptstyle
N\Delta}\}_{sr}$
\begin{eqnarray}
\{\Psi_{\scriptscriptstyle NN}\}_{sr}&=&
\sum_f C^{\scriptscriptstyle (N)}_{\scriptscriptstyle f}\,d_f(s^4p^2)=
{\cal A}\,\{\varphi_{\scriptscriptstyle 2S}(r)N(1,2,3)N(4,5,6)\}
\ \ \mathrm{and}\nonumber\\
\{\Psi_{\scriptscriptstyle N\Delta}\}_{sr}&=&
\sum_f C^{\scriptscriptstyle (\Delta)}_{\scriptscriptstyle f}\,
D_f(s^4p^2)={\cal A}\,
\{\varphi_{\scriptscriptstyle 2S}(r)N(1,2,3)\Delta(4,5,6)\},
\label{sup}
\end{eqnarray}
where ${\cal A}$ is the antisymmetrizer 
${\cal A}=\frac{1}{10}(I-\sum_{i=1}^{3}\sum_{j=4}^{6}P_{ij})$ and 
$N(1,2,3)$ and $\Delta(4,5,6)$ are the nucleon and $\Delta$-isobar states with
fixed numbers of quarks $i$($j$).  The coefficients
 $C^{\scriptscriptstyle (N)}_{\scriptscriptstyle f}$ and
 $C^{\scriptscriptstyle (\Delta)}_{\scriptscriptstyle f}$ are algebraic ones
apart from a common energy-dependent factor (which can be calculated in
principle by solving the RGM equations, see e.g. Ref.~\cite{kus}).  
It is particularly remarkable that these cluster-like states correspond 
to a nodal wave function~$\varphi_{\scriptscriptstyle 2S}(r)$ which describe 
the relative motion of two clusters with 
 ${\bf r}=\frac{1}{3}({\bf r}_1+{\bf r}_2+ {\bf
 r}_3-{\bf r}_4-{\bf r}_5-{\bf r}_6)$ as the relative coordinate.  In the
 present case it is the $2S$~state of the h.o. basis
 $\varphi_{\scriptscriptstyle 2S}(r)\sim(1-{r^2}/{b^2})\,e^{-3r^2/4b^2}$, but
 in general it would be a more realistic nodal wave function. Here $b$ is the
 h.o. radius, which is a scale parameter with a typical value of
 $b\approx\,$0.5-0.6~fm.

Such a decomposition of the total antisymmetrized short-range $NN$ ($N\Delta$)
 wave function into two mutually orthogonal parts:

\noindent
 (i) a non-nucleonic (bag-like) state
\begin{eqnarray}
d_0(s^6)=|s^6[6]_{\scriptscriptstyle X}\,[3^2]_{\scriptscriptstyle ST}\,
(ST=10),L{=}0\,\rangle\quad
\mbox{(for the $NN$ overlap)},\nonumber\\
D_0(s^6)=|s^6[6]_{\scriptscriptstyle X}\,[3^2]_{\scriptscriptstyle ST}\,
(ST=21),L{=}0\,\rangle\quad
\mbox{(for the $N\Delta$ overlap)},
\label{dD}
\end{eqnarray}

\noindent
 and (ii) a ``proper'' $NN$ (or $N\Delta$) component [Eq.~(\ref{sup}) with 
 $f\ne 0$], is
 characteristic of any quark-model approach. This property is independent
 on the choice of the quark-quark interaction. It is  a specific 
 manifestation of the Pauli exclusion principle in the system of six 
 colored quarks.  
 It is interesting that in the $ST=11$ $N\Delta$ channel
 a symmetric bag-like state~$s^6$ of Eq.~(\ref{dD})-type is
 forbidden, and thus the complete wave function of the $N\Delta$ system 
 at short ranges in that channel  reduces to a net nodal cluster-like 
 state of Eq.~(\ref{sup})-type. In a general case the $NN$ or $N\Delta$
 wave function should be written as a two-component Fock column.  

\begin{figure}[hp]
\begin{center}
\epsfig{file=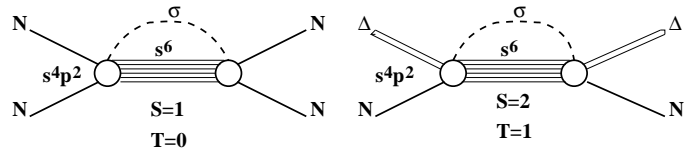,width=0.5\textwidth,clip}
\caption{\label{f3}
 Loop diagram contributions to the short-range $NN$ and $N\Delta$ interaction.
}
\end{center}
\end{figure}

To be precise, the representation (Eq.\ref{sup}) of the cluster wave function
 $\Psi_{\scriptscriptstyle NN}(r)$ or $\Psi_{\scriptscriptstyle N\Delta}(r)$
 is only meaningful in the region $0\leqslant r \lesssim b\div 2b$.  The full
 wave function of the system has a two-channel structure, namely the
 two-component Fock column:
\begin{equation}
{\Psi_{\scriptscriptstyle NN}\choose{d_0(s^6)+\mbox{cloud}}}
\quad{\mbox or}\quad
{\Psi_{\scriptscriptstyle N\Delta}\choose{D_0(s^6)+\mbox{cloud}}},
\label{fock}
\end{equation}
In Fig.~\ref{f3}, the ``cloud'' is represented by the $\sigma$-meson
 $s$-channel loop.

A similar picture can also be observed in the lowest odd partial wave,
 i.e. $L=1$, where the configuration $s^5p[51]_{\scriptscriptstyle X}$ has a
 trivial symmetry $[51]_{\scriptscriptstyle X}$ in coordinate space and its
 spin-isospin wave function is strictly constrained by the Pauli exclusion
 principle. The corresponding bag-like state in the channel $ST=11$
\begin{equation}
d_1(s^5p)=|s^5p[51]_{\scriptscriptstyle X}[321]_{\scriptscriptstyle ST}
(S_1T_1=11),L_1{=}1\,\rangle
\label{d1}
\end{equation}
is also contributing to the considered processes. At the same time, in the
 excited configuration $s^3p^3[3^2]_{\scriptscriptstyle X}$ the ``true''
 cluster-like $NN$ (or $N\Delta$) states are allowed, and a corresponding
 cluster wave function $\varphi_{\scriptscriptstyle 3P}(r)$ has also a node at
 short ranges.

However, for the total angular momentum $J^P=2^+$ of the initial $N\Delta$
state a full set of possible $d_1+\pi$ intermediate states with the total
isospin $T=1$ includes 11 shell-model configurations 
$d_1(s^5p)$ with different values of the spin $S_1=1,2,3$ (in the simplest
case of the $S$-wave pion), the
isospin $T_1=0,1,2$ and the Young tableaux 
$[f_{\scriptscriptstyle ST}]=[321],\,[42]$. The configuration~(\ref{d1})
is only an example of the basis vector from this set.

\subsection{Dressed-bag model}
\label{ssec:dbmodel}

It has been proposed~\cite{kuk}, that the non-nucleonic bag-like states
 $s^6[6]_{\scriptscriptstyle X}$ and $s^5p[51]_{\scriptscriptstyle X}$ with
 trivial coordinate symmetry in two lowest partial waves $L=$~0 and~1,
 respectively, (in $NN$ and $N\Delta$ channels), acquire their own meson
 clouds which are different from the sum of two meson clouds of the initial
 baryons. Probably, such a cloud will serve to stabilize the six-quark bag in
 the $NN$ (or $N\Delta$) overlap region. For example, it looks tempting to
 consider the scalar-isoscalar ($\sigma$-meson) field as most strong in the
 presence of the six-quark bag. The quark masses are reduced in the scalar
 field, and such a ``dressed'' bag (see Fig.~\ref{f3}) should be much lower in
 energy than the corresponding six-quark shell-model states $s^6$ ($s^5p$).

Thus we come to the following dynamical model~\cite{kuk} which can be
 tested in numerous processes:

\begin{itemize}
\item[(i)] At short ranges there is an enhanced $\sigma$-field originating
   from transitions between the ``proper'' $NN$ cluster-like state
   $\{\Psi_{\scriptscriptstyle NN}\}_{sr}$ and the
   bag-like one  $d_0(s^6)$
\begin{equation}
	\{\Psi_{\scriptscriptstyle NN}\}_{sr}\leftrightarrows d_0+\sigma
\label{tr}
\end{equation}
\item[(ii)] The short-range $NN$ dynamics is determined by
	coupled-channel equations for the two channels of Eq.~(\ref{tr}).
\item[(iii)] An effective separable and energy-dependent $NN$ potential
   $V_{NqN}(r,r^{\prime};E)$ emerges after exclusion of the closed channel
   $d_0+\sigma$ by use of  standard methods~\cite{fesh} of nuclear physics.
   This potential is related to the corresponding $s$-channel loop diagram (see
   Fig.~\ref{f3}), the contribution of which to the short-range $NN$ force can
   be calculated in terms of the $^3P_0$ model (see Ref.~\cite{kuk}).
\end{itemize}

The effective potential for the short-range $N\Delta$ interaction is
 determined by the same scheme of calculations.  Since the method has been
 described in Ref.~\cite{kuk} we sketch here only the basic outline of the
 calculation.

\begin{figure}[hp]
\begin{center}
\epsfig{file=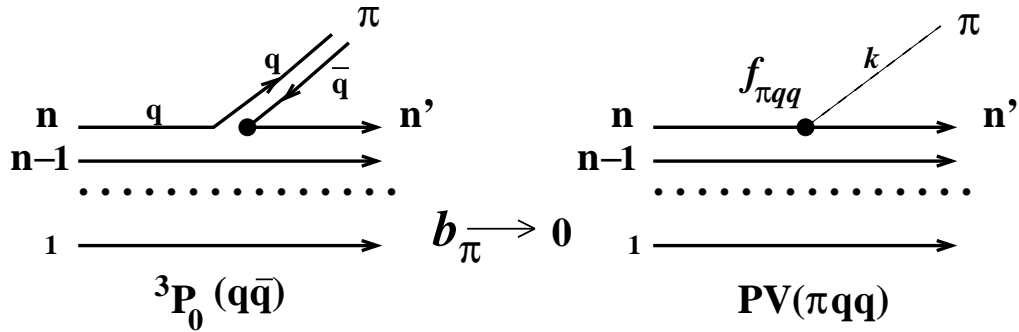,width=0.75\textwidth,clip}
\caption{\label{f4}
Quark-pion coupling in the $n$-quark system.}
\end{center}
\end{figure}

We start from the Hamiltonian for quark-pion coupling deduced from the $^3P_0$
 model~\cite{yao} in the form that takes into account the Galilean invariance
 for the system of $n$ constituent quarks (n=3 or 6, see numbering in
 Fig.~\ref{f4}) with equal masses $m_q=m_N/3$ (see the details
 in Refs.~\cite{kuk,obf,can}),
\begin{eqnarray}
H^{(n)}_{\pi qq}({\bf k},\lambda)&=&-i\,\frac{f_{\pi qq}}{m_\pi}\,
\tau_{\scriptscriptstyle-\lambda}
e^{i\frac{n-1}{n}{\bf k\cdot}{\b\rho}}\,O({\b\rho},{\b\rho^{\prime}})\,
{\b\sigma}\cdot\left[\frac{\omega_{\pi}}{2m_q}
(\frac{2}{i}{\b\nabla}_{\rho}+\frac{n-1}{n}{\bf k})
+(1+\frac{\omega_{\pi}}{2nm_q}){\bf k}\right]\ \ \mathrm{and}\nonumber\\
O({\b\rho},{\b\rho^{\prime}})&=&
e^{-i\frac{1}{2}{\bf k\cdot({\b\rho\!-\!\b\rho^{\prime}})}}
\Psi_{\pi}({\b\rho}\!-\!{\b\rho^{\prime}}),
\label{piqq}
\end{eqnarray}
written here in relative coordinates of the $n$-th quark
 ${\b\rho}=\frac{1}{n-1}\sum_{i=1}^{n-1}{\bf r}_i\!-\!{\bf r}_{n}$.  It is
 coincident with a local pseudo-vector (PV) coupling (taken in a
 non-relativistic approximation), when the pion radius $b_{\pi}$ approaches
 zero (Fig.~\ref{f4})
\begin{equation}
\Psi_{\pi}({\b\rho}\!-\!{\b\rho^{\prime}})\to
\delta({\b\rho}\!-\!{\b\rho^{\prime}})\ \ \mathrm{for}\ \ \quad b_{\pi}\to 0.
\label{lim}
\end{equation}

\begin{figure}[htbp]
\begin{center}
\epsfig{file=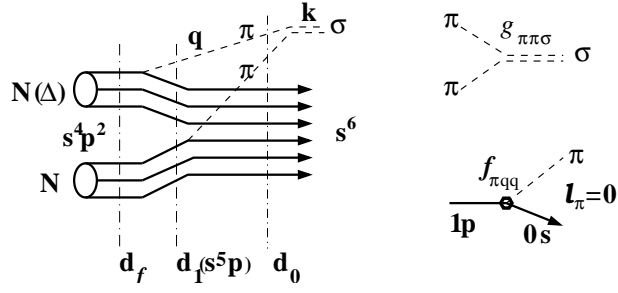,width=0.45\textwidth,clip}
\caption{\label{f5}
Triangle loop diagram of $\sigma$-meson emission
and its vertices.}
\end{center}
\end{figure}

The $\sigma$-meson is considered as a broad resonance ($\Gamma_{\sigma}\approx
 m_{\sigma}\approx\,$500~MeV) in the $\pi$-$\pi$ system with a simple vertex
 function
\begin{equation}
v_{\pi\pi\sigma}({\bf k},{\bf k^{\prime}})\equiv
\langle\,\sigma({\bf k}+{\bf k}^{\prime})|
H_{\sigma\pi\pi}|\,\pi({\bf k})\pi({\bf k}^{\prime})\,\rangle=
g_{\pi\pi\sigma}F_{\pi\pi\sigma}(q^2)
\label{sigv}
\end{equation}
with ${\bf q }=({\bf k}\!-\!{\bf k}^{\prime})/2$ and with a Gaussian form
 factor $F_{\pi\pi\sigma}(q^2)=e^{-q^2b_{\sigma}^2/2}$ and a standard value
 $g_{\pi\pi\sigma}\approx\,$2-4~GeV~\cite{sch}.  The $\sigma$ is emitted from
 the cluster-like configuration $s^4p^2$ in transition to the bag-like $s^6$
 through two subsequent pion emissions with the de-excitation of p-shell quarks
 $d_f(s^4p^2)\to d_1(s^5p)\to d_0(s^6)$ (or $D_f(s^4p^2)\to D_1(s^5p)\to
 D_0(s^6)$) as is shown in Fig.~\ref{f5}.

Starting from the representation (\ref{sup}) of a short-range $NN$ (or
$N\Delta$) wave function, the $NN\to d_0+\sigma$ (or $N\Delta\to D_0+\sigma$)
vertex function $\Gamma_{\scriptscriptstyle N\sigma q}$ (or
$\Gamma_{\scriptscriptstyle \Delta\sigma q}$) can be written in the form
\begin{equation}
\Gamma_{\scriptscriptstyle N\sigma q}(k,E)=
\sum_fC^{(\scriptscriptstyle N)}_f
	\Gamma^{(\scriptscriptstyle N)}_f(k,E),\qquad
	\qquad \Gamma_{\scriptscriptstyle \Delta\sigma q}(k,E)=
\sum_fC^{(\scriptscriptstyle \Delta)}_f
	\Gamma^{(\scriptscriptstyle \Delta)}_f(k,E),
\label{vert}
\end{equation}
where the partial vertices $\Gamma^{(\scriptscriptstyle N,\,\Delta)}_f(k,E)$
  are quark-shell-model matrix elements for the  transitions $d_f\to d_0+\sigma$
  ($D_f\to D_0+\sigma$) shown in Fig.~\ref{f5}:
\begin{eqnarray}
\Gamma^{(\scriptscriptstyle N)}_f(k,E)= 15\times\sum_{\lambda}
	\left(1\lambda 1-\!\lambda|00\right)
\sum_{d_1}\int d^3q&&\left\langle
\sigma({\bf k})|H_{\sigma\pi\pi}|\pi({\bf q})\pi({\bf q\!-\!k})\right\rangle
\frac{1}{E-H_{2\pi\!+\!d_0}}\nonumber\\
&&\times\left\langle d_0|H^{(5)}_{\pi qq}({\bf q\!-\!k},-\lambda)|d_1\right\rangle
\frac{1}{E-H_{\pi\!+\!d_1}}
\left\langle d_1|H^{(6)}_{\pi qq}({\bf q},\lambda)|d_f\right\rangle.
\label{parv}
\end{eqnarray}

Here $H_{2\pi\!+\!d_0}$ and $H_{\pi\!+\!d_1}$ are undisturbed energies of the
 intermediate states $2\pi+d_0$ and $\pi+d_1$ of the transitions
\begin{equation}
H_{\pi\!+\!d_1}=m_{d_0}+\frac{k^2}{2m_{d_0}}+\omega_{\pi}({\bf q})+
\omega_{\pi}({\bf q}-{\bf k})\ \ \mathrm{and}\ \qquad
H_{\pi\!+\!d_1}=m_{d_1}+\frac{k^2}{2m_{d_1}}+\omega_{\pi}({\bf q})
\label{kin}
\end{equation}
with $m_{d_0}\approx m_{d_1}\approx\,$2300-2500~MeV as a phenomenological
 parameter. In Eq.~(\ref{parv}) the sum over $d_1$ includes all the 11 
$d1(s^5p)$ configurations listed after Eq.~(\ref{d1}) in the previous
Subsection. 

The spin- and isospin-flip matrix elements of the operators defined in
 Eq.~(\ref{piqq}) for 5-th and 6-th quarks $\langle d_0|H^{(5)}_{\pi qq}({\bf
 q\!-\!k},-\lambda)|d_1\rangle$ and $\langle d_1|H^{(6)}_{\pi qq}({\bf
 q},\lambda)|d_f\rangle$ are calculated in the limit of Eq.~(\ref{lim}) with
 the standard fractional parentage coefficients (f.p.c.) technique of the
 quark shell model (see refs.~\cite{kuk,har,obu,obf} for details).  The results
 for $NN$~\cite{kuk} and $N\Delta$ channels are similar to each other:
\begin{equation}
\Gamma_{\scriptscriptstyle N\sigma q}(k,E)=
\frac{f_{\pi qq}^2}{m_{\pi}^2}\,\frac{g_{\sigma\pi\pi}}{m_q^2b^3}\,
C^{\scriptscriptstyle N}_{\scriptscriptstyle ST}\,{\cal D}(k,E)
\ \ \mathrm{and}\ \qquad
\Gamma_{\scriptscriptstyle \Delta\sigma q}(k,E)=
\frac{f_{\pi qq}^2}{m_{\pi}^2}\,\frac{g_{\sigma\pi\pi}}{m_q^2b^3}\,
C^{\scriptscriptstyle \Delta}_{\scriptscriptstyle ST}\,{\cal D}(k,E)\ .
\label{vt}
\end{equation}
Here only the algebraic factors $C^{\scriptscriptstyle N}_{\scriptscriptstyle
10}=\frac{-19}{486}\sqrt{\frac{5}{3}}$ and $C^{\scriptscriptstyle
\Delta}_{\scriptscriptstyle 21}=\frac{113}{3645}\sqrt{\frac{5}{6}}$ 
are different
for $NN$ and $N\Delta$ channels because of different spin-isospin quantum
numbers $ST=$~10 and~21. Hence the effective ``coupling constants'' for
transitions of Eq.~(\ref{tr}) can be defined in a simple form dependent on
fixed parameters of the model ($\pi qq$ and $\sigma\pi\pi$ constants,
masses~$m_{\pi}$, $m_{\sigma}$, h.o. radius $b$) and an algebraic factor
depending on the $ST$ values
\begin{equation}
g_{\scriptscriptstyle N\sigma q}=g_{\scriptscriptstyle \sigma qN}=
\frac{f_{\pi qq}^2}{m_{\pi}^2}\,\frac{g_{\sigma\pi\pi}}{m_q^2b^3}\,
C^{\scriptscriptstyle N}_{\scriptscriptstyle ST}\ \ \mathrm{and}\ \qquad
g_{\scriptscriptstyle \Delta\sigma q}=
g_{\scriptscriptstyle \sigma q\Delta}=
\frac{f_{\pi qq}^2}{m_{\pi}^2}\,\frac{g_{\sigma\pi\pi}}{m_q^2b^3}\,
C^{\scriptscriptstyle \Delta}_{\scriptscriptstyle ST}
\label{g}
\end{equation}

\begin{figure}[htb]
\begin{center}
\epsfig{file=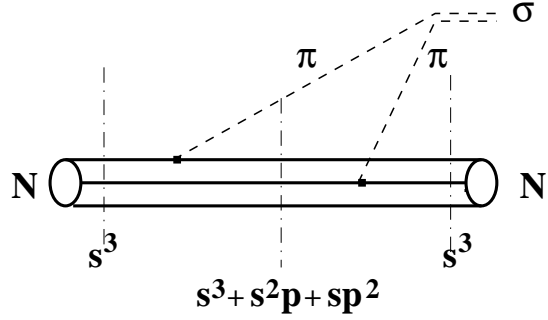,width=0.4\textwidth,clip}
\caption{\label{f6}
Triangle loop diagram for the $\sigma$-nucleon coupling
		constant $g_{\scriptscriptstyle \sigma NN}$.}
\end{center}
\end{figure}

However, the ``form factor'' ${\cal D}(k,E)$ (common for both above channels
 as well) has a more uncertain magnitude since it is obtained by the
 integration over the inner momenta~$q$ of the triangle loop diagram
 (Fig.~\ref{f5}), and thus it depends on the generally unknown high-momentum
 behaviour of the $\pi qq$ and $\sigma\pi\pi$ form factors.  After a simple
 modification of each CQM Gaussian in the form
\begin{equation}
e^{-\alpha^2 q^2}\to\,\, e_{mod}(-\alpha^2 q^2)=(1-p)\,e^{-\alpha^2 q^2}+
p\,e^{-\alpha^2 q^2/z^2},
 \ \mathrm{with}\ \ p=\frac{1}{3}\ \mathrm{and} \quad
z\approx 2,
\label{mod}
\end{equation}
which does not dramatically change the low-momentum behaviour of all the CQM
form factors, but enhances its high-momentum component by modifying the
h.o. scale $b\to b/z$~%
\footnote{The h.o. radius $b\approx$0.5-0.6~fm of the non-relativistic quark
          model corresponds to the characteristic confinement scale of the QCD
          $\Lambda_{QCD}\approx$200~MeV, but the high-momentum behaviour of
          $F_{QM}$ in loop integrals should also depend on a higher
          characteristic scale $\Lambda_{ch}\approx$0.6-1~GeV related to the
          spontaneous breaking of chiral symmetry.}%
, the contribution of the diagram in Fig.~\ref{f6} to the $\sigma NN$ coupling
 leads to a realistic value $g_{\scriptscriptstyle \sigma NN}$. It should be
 noted that in contrast to a simple version proposed in Ref.~\cite{gut} our 
 calculations imply the summation over all possible three-quark intermediate 
configurations up to 2 h.o. quanta.

\subsection{Reduction to effective non-local $NN$ and $N\Delta$
interactions}
\label{ssec:effint}

\begin{figure}[hbtp]
\begin{center}
\epsfig{file=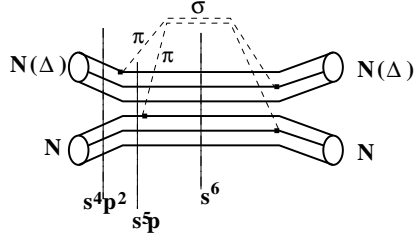,width=0.3\textwidth,clip}
\caption{\label{f7}
The s-channel loop diagram of $NN$ ($N\Delta$)
interaction.}
\end{center}
\end{figure}

The $s$-channel loop diagram in Fig.~\ref{f3} with vertices calculated in 
the previous subsection (see Fig.~\ref{f7})
 can be reduced to a non-local attractive $NN$
(or $N\Delta$) interaction potential of the form
\begin{equation}
\label{vnn}
V_{NqN}(r,r^{\prime};E)=\int d^3k\,\varphi_{2s}(r^{\prime})\,
\Gamma_{\scriptscriptstyle N\sigma q}(k,E)\frac{1}{E-H_{\sigma+d_0}}
\Gamma_{\scriptscriptstyle q\sigma N}(k,E)
\varphi_{2s}(r)
\end{equation}
After integration over the inner momentum {\bf k} the following expressions
for effective $NN$ and $N\Delta$ potentials arise
\begin{equation}
V_{\scriptscriptstyle NqN}(r,r^{\prime};E)=
g_{\scriptscriptstyle N\sigma q}^2
I(E)|\,\varphi_{\scriptscriptstyle 2S}(r^{\prime})\,\rangle\,
\langle\,\varphi_{\scriptscriptstyle 2S}(r)\,|\ \ \mathrm{and}\ \quad
V_{\scriptscriptstyle \Delta q\Delta}(r,r^{\prime};E)=
g_{\scriptscriptstyle \Delta\sigma q}^2
I(E)|\,\varphi_{\scriptscriptstyle 2S}(r^{\prime})\,\rangle\,
\langle\,\varphi_{\scriptscriptstyle 2S}(r)\,|,
\label{eff}
\end{equation}
where the energy-dependent function $I(E)$ is common for both potentials
\begin{equation}
I(E)=\int d^3k\,\frac{{\cal D}(k,E)^2}{E-H_{\sigma+d_0}}
\label{lam}
\end{equation}
Note that such a factorization of $V_{NqN}(r,r^{\prime};E)$ and $V_{\Delta
 q\Delta}(r,r^{\prime};E)$ into an ``algebraic part'' (constants
 $g_{\scriptscriptstyle N\sigma q}^2$ and $g_{\scriptscriptstyle \Delta\sigma
 q}^2$) and a common for both channels integral part $I(E)$ is
 characteristic of the dressed-bag model.

The significance of this effective interaction is two-fold:
\begin{itemize}
\item[({i})] it projects the short-range $NN$ (or $N\Delta$ ) wave function
    onto the nodal $2S$ state through the projection operator
\begin{equation}
\hat P_{\scriptscriptstyle 2S}=
|\,\varphi_{\scriptscriptstyle 2S}(r^{\prime})\,\rangle
\langle\,\varphi_{\scriptscriptstyle 2S}(r)\,|,
\label{pr}
\end{equation}
  which determines a specific form of the short-range correlations (e.g. these
  correlations are seen in the $NN$ scattering as a ``repulsive-core-like''
  behaviour of the S-wave phase shift \cite{kuk});

\item[({ii})] it emphasizes the Pauli exclusion principle which imposes
the restriction not only on the form but also on the
relative strength [$g_{\scriptscriptstyle N\sigma q}$ and
$g_{\scriptscriptstyle \Delta\sigma q}$ in Eq.~(\ref{g})] of the
short-range interaction in each $ST$ channel.
\end{itemize}

To gain a better understanding of the proposed mechanism for the short-range
 interaction an analytical (Pade-) approximation for the integral in
 Eq.~(\ref{lam}) as a function of energy E has been performed in
 Ref.~\cite{kuk}. An accurate approximation for $I(E)$ exhibits a pole
 behaviour
\begin{equation}
I(E)\approx \lambda_0\,\frac{E_0-a_0E}{E_0-E},
\label{pade}
\end{equation}
with $E_0$ close to the mass of the primitive state $s^6+\sigma$. The value of
 $a_0$ is too small ($a_0\approx0.05$) to lead to a detectable modification of
 the pole behaviour of Eq.~(\ref{pade}).  But the value $\lambda_0$ calculated
 with the same set of parameters as it was used in the above calculation of
 the coupling constant $g_{\sigma NN}$ turns out too small to give a realistic
 description of the $NN$ phase shifts. We found that a considerable 
 renormalization
 of $\lambda_0$ (factor 2 to 4) is necessary~\cite{kuk} to obtain a good
 description for the $NN$ phase shifts up to 1~GeV.  However, the considered
 mechanism is qualitatively correct.

The quantitative description is also possible~\cite{kuk} by fitting the values
 $\lambda_0$, $E_0$ and $a_0$ as free parameters in order to describe the $NN$
 phase shift. In such a procedure the $NN$ interaction through meson exchange
 is also considered for long ranges $r\gtrsim b$.  The following parameters
 for the short-range $NN$ interaction were obtained \cite{kuk}:
\begin{eqnarray}
g_{\scriptscriptstyle N\sigma q}^2\,\lambda_0=
328.55\,\,\mbox{MeV}, \quad E_0-2m_N=693\,\,\mbox{MeV} \quad
a_0=-0.05\ \ \mathrm{and}\ \ \quad  b=0.5065\,\,\mbox{fm}
\label{param}
\end{eqnarray}
which are used in the present work to describe the $N\Delta$ short-range
 interaction in the $ST=21$ channel with the substitution
 $g_{\scriptscriptstyle N\sigma q}\to g_{\scriptscriptstyle \Delta\sigma q}$.
Such substitution implies that effective coupling constant 
$g_{\scriptscriptstyle N\sigma q}$ should be roughly reduced in 1.7 
times for $N\Delta$ interaction,  and the effective parameter $\lambda_0$ in 
the separable potential (\ref{eff})-(\ref{pade}) should be roughly reduced 
in three times. But such evaluations of the interaction on the basis of 
algebraic factors in Eq.~(\ref{vt}) are only reliable to the order of 
magnitude, and thus our calculations of the effect of $N\Delta$ FSI have
been performed for the reduced and non-reduced value of the constant 
$\lambda_0$ (see Sect.~\ref{sec:deuteron}).

\subsection{The $N\Delta$ and ``dibaryon'' mechanisms of pion emission
from the deuteron}
\label{ssec:srinter}
 
\begin{figure}[hp]
\begin{center}
\epsfig{file=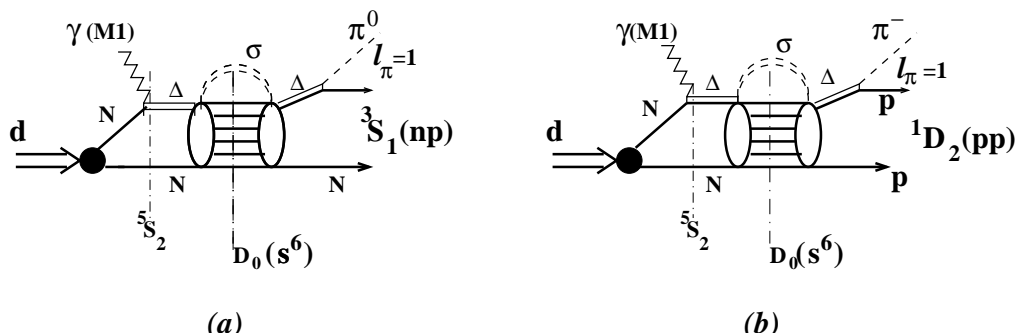,width=0.75\textwidth,clip=}
\caption{\label{f8}
The FSI of $N\Delta$ type with a $P$-wave pion.
}
\end{center}
\end{figure}

\begin{figure}[hbtp]
\begin{center}
\epsfig{file=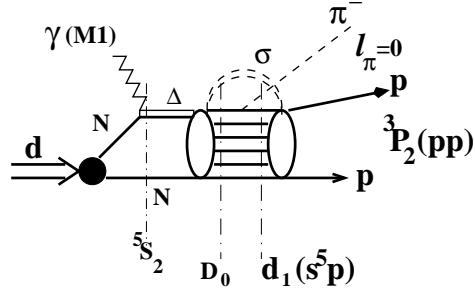,width=0.35\textwidth,clip=}
\caption{\label{f9}
The FSI of ``dibaryon'' type with the $S$-wave pion.}
\end{center}
\end{figure}

In the $\pi^\mathrm{o}$ and $\pi^-$ photoproduction off the deuteron there are
at least two different mechanisms of pion emission shown in
 Figs.~\ref{f8} and~\ref{f9}.
In both cases there is the same intermediate $N\Delta$ state 
 in the $^5S_2$ partial wave generated
 by the isovector $M1$ $\gamma N\to\Delta$ transition on one of the nucleons.
 Note that for the other $N\Delta$ partial wave $^3S_1$ the analogous bag-like
 component $D_0(s^6)$ with $ST=11$ is forbidden. However the final
 $\pi^\mathrm{o}pn$ and $\pi^-pp$ states are different in isospin of the $NN$
 subsystem.  As a result the lowest partial wave $^3S_1$ in the $np$ final
 state is allowed, but in the final $pp$ state ($T$=1) the lowest $^1S_0$ wave
 is forbidden.  The $^1D_2(pp)$ partial wave is only allowed through
 mechanisms shown in Fig.~\ref{f8}(b).

However, the $D_0(s^6)$ bag-like state with $ST=21$ can bypass the transition
 to the final $^5S_2(N\Delta)$ state allowing the direct transition to the
 final $\pi^-pp$ state with a $^3P_2(pp)$ partial wave with the $S$-wave pion,
 as it is shown in Fig.~\ref{f9}.  In this case the $P$-wave bag-like state
 $d_1(s^5p)$ of Eq.~(\ref{d1}) is produced after the $S$-wave $\pi^\mathrm{o}$
 emission instead of the even-parity final state $^1D_2(pp)$ produced by
 $P$-wave $\pi^-$ emission.  Hence the dressed bag $d_1(s^5p)+\sigma$ should
 contribute to the $\pi^-$ emission and could be seen as a correlated $pp$
 pair in the $^3P_2-^3F_2$ partial waves.  Such a correlated $pp$ pair has to
 absorb the recoil pion momentum as a whole. A possible dibaryon resonance
 with the quantum numbers of $d_1(s^5p)+\sigma$ system could enhance the role
 of such a correlated $pp$ pair, and thus any polarization experiments which
 will be able to make a selection of the final $pp$ pair in the $^3P_2-^3F_2$
 states would be desirable.

But in general (i.e. without the dibaryon enhancement) the role of $N\Delta$
 FSI interaction in the $\pi^-$ photoproduction should be considerably smaller
 than in the $\pi^\mathrm{o}$ case. In Sect.~\ref{sec:deuteron} this
 qualitative analysis will be supported by quantitative evaluations made in
 the framework of the CQM.

\section{Pion photoproduction off the nucleon}
\label{sec:nucleon}

First we want to test our approach in pion photoproduction from free
 nucleons. These processes are enhanced in the resonance region by the
 excitation of virtual baryons.  In the CQM, as a rule, we deal with bare
 (i.e. three-quark) baryons which can be treated as h.o. 3q configurations in
 the translationally-invariant shell model (TISM)
\begin{eqnarray}
N(940)=|s^3[3]_{\scriptscriptstyle X}L\!&=&
\!0\rangle_{\scriptscriptstyle TISM}\,
\cdot\,|\,[1^3]_{\scriptscriptstyle C}[3]_{\scriptscriptstyle ST}
S=1/2,\,T=1/2\rangle,\nonumber\\
\Delta(1232)=|s^3[3]_{\scriptscriptstyle X}L\!&=&\!0
\rangle_{\scriptscriptstyle TISM}\,
\cdot\,|\,[1^3]_{\scriptscriptstyle C}[3]_{\scriptscriptstyle ST}
S=3/2,\,T=3/2\rangle,\nonumber\\
R(1440)=|sp^2[3]_{\scriptscriptstyle X}L\!&=&\!0
\rangle_{\scriptscriptstyle TISM}\,
\cdot\,|\,[1^3]_{\scriptscriptstyle C}[3]_{\scriptscriptstyle ST}
S=1/2,\,T=1/2\rangle\ \ \mathrm{and}\nonumber\\
N^*(1535)=|s^2p[21]_{\scriptscriptstyle X}L\!&=&\!1
\rangle_{\scriptscriptstyle TISM}\,
\cdot\,|\,[1^3]_{\scriptscriptstyle C}[3]_{\scriptscriptstyle ST}
S=1/2,\,T=1/2\rangle
\label{bar}
\end{eqnarray}
with a simple structure of the coordinate part, e.g.
\begin{eqnarray}
|\,s^3[3]_XL&=&0\rangle_{\scriptscriptstyle{TISM}}=
\varphi_{0S}(\rho_1/\alpha_1)\varphi_{0S}(\rho_2/\alpha_2),\\
|\,sp^2[3]_XL&=&0\rangle_{\scriptscriptstyle{TISM}}=
\sqrt{\frac{1}{2}}\,
[\varphi_{0S}(\rho_1/\alpha_1)\varphi_{2S}(\rho_2/\alpha_2)+
\varphi_{2S}(\rho_1/\alpha_1)\varphi_{0S}(\rho_2/\alpha_2)] , \dots ,etc.,
\end{eqnarray}
where the h.o. basis wave functions
\begin{equation}
\varphi_{0S}(\frac{{\b\rho}}{\alpha})=
(\pi\alpha^2)^{-3/4}e^{-{\b\rho}^2/2\alpha^2},\quad
\varphi^m_{1P}(\frac{{\b\rho}}{\alpha})=
(\pi\alpha^2)^{-3/4}\sqrt{\frac{2}{3}}\frac{\rho}{\alpha}
e^{-{\b\rho}^2/2\alpha^2}\sqrt{4\pi}Y_{1m}(\hat{\b\rho}),\dots, etc.
\end{equation}
depend only on the relative quark coordinates
${\b\rho}_1={\bf r}_1-{\bf r}_2$, ${\b\rho}_2=({\bf r}_1+{\bf r}_2)
/2-{\bf r}_3$, $\alpha_1=\sqrt{2}\,b$, $\alpha_2=\sqrt{3/2}\,b$.

The most important ingredient of any model for photoproduction is the
 procedure for the dressing of the bare baryons by meson fields. Here we
 consider a procedure starting from the $^3P_0$ model~\cite{yao,obf,can}.

\subsection{Dressing of bare baryons and the off-shell $\Delta$
propagator}
\label{ssec:baryonstates}

\begin{figure}[hbtp]
\begin{center}
\epsfig{file=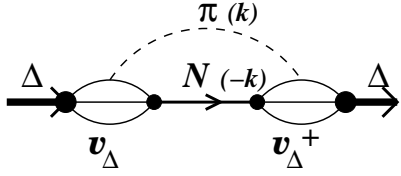,width=0.3\textwidth,clip=}
\caption{\label{f10}
 Dressing of the $\Delta$-isobar.}
\end{center}
\end{figure}

The bare baryons should be dressed by the pion field to acquire finite decay
 widths. Then, for example, the $\Delta$ isobar acquires a finite decay width
 $\Gamma_{\Delta}\approx120$~MeV, but the nucleon remains stable.  For
 simplicity all the calculations are performed with the undressed w.f. for the
 nucleon, but for the $\Delta$-isobar a standard dressing procedure (see,
 e.g.~\cite{kum}) based on a self-energy loop diagram contribution to the mass
 of $\Delta$ (see diagram in Fig.~\ref{f10}) has been performed:
\begin{equation}
\label{sig}
\Sigma_{\Delta}(E)=\sum_{\lambda}
\int\frac{d^3k}{(2\pi)^32\omega_{\pi}}\,
\frac{v^{\dag}_{\Delta}({\bf k},\lambda) v_{\Delta}({\bf k},\lambda)}
{E-M_N-k^2/2M_N-\omega_{\pi}(k)+i0}
\end{equation}
The $\pi N\Delta$ vertex function $v_{\Delta}({\bf k},\lambda)$ in
 Eq.~(\ref{sig}), where ${\bf k}$ is a (relative) momentum of the pion
 $\pi^{\lambda}$, $\lambda=0,\,\pm 1$ in the $\pi N$ system, has been
 calculated within the CQM using the $^3P_0$ model as a matrix element of the
 operator Eq.(\ref{piqq}) for the transition $\Delta\to \pi N$ in the limit of
 Eq.(\ref{lim}).
\begin{equation}
\label{vd}
v_{\Delta}({\bf k},\lambda)=
3\langle N(\mu_{\scriptscriptstyle N},
t_{\scriptscriptstyle N})|H^{(3)}_{\pi qq}({\bf k},\lambda)|
\Delta(\mu_{\scriptscriptstyle \Delta},
t_{\scriptscriptstyle \Delta})\rangle=
i\left(\frac{2\sqrt{2}
f_{\pi qq}}{m_{\pi}}\right)
\,e^{-k^2b^2/6}\,{\bf S}\cdot{\bf k}\,T_{\lambda},
\end{equation}

In Eq. (\ref{vd}) the transition operators $S_m$($T_{\lambda}$) are components
 of the total spin (isospin) of the $3q$ system in the spherical basis. Their
 matrix elements related to the spin (isospin) Clebsch-Gordan coefficients
 through the standard formulas
\begin{equation}
\langle\frac{1}{2}\mu_{\scriptscriptstyle N}|S_m|
\frac{3}{2}\mu_{\scriptscriptstyle \Delta}\rangle=
\sqrt{2}\,(\frac{3}{2}\,\mu_{\scriptscriptstyle \Delta}\,1\,m\,|\,
\frac{1}{2}\,\mu_{\scriptscriptstyle N})\ \ \mathrm{and}\ \ \quad
\langle\frac{1}{2}t_{\scriptscriptstyle N}|T_{\lambda}|
\frac{3}{2}t_{\scriptscriptstyle \Delta}\rangle=
\sqrt{2}\,(\frac{3}{2}\,t_{\scriptscriptstyle \Delta}\,1\,\lambda\,|\,
\frac{1}{2}\,t_{\scriptscriptstyle N}).
\end{equation}
Here the notations $\mu_{\scriptscriptstyle \Delta}$($t_{\scriptscriptstyle
 \Delta}$) and $\mu_{\scriptscriptstyle N}$($t_{\scriptscriptstyle N}$) are
 used for the spin (isospin) projections of the initial $\Delta$-isobar and
 the intermediate nucleon $N$ as it is shown in the left-hand side of the
 diagram in Fig.~\ref{f10}.

The self energy (\ref{sig}) modifies the bare mass $M^0_{\Delta}(3q)$
 of the three-quark shell-model ${\Delta}$-isobar state
\begin{equation}
M_R(E)=M^0_{\Delta}(3q)+{\cal R}e\Sigma_{\Delta}(E)
\label{mr}
\end{equation}
and the imaginary part of $\Sigma_{\Delta}(E)$ defines the $\Delta$-isobar
 width
\begin{equation}
\Gamma_R(E)=-2\ {\cal I}m\Sigma_{\Delta}(E)
\label{gr}
\end{equation}
Due to the self-energy part of the mass operator the free $\Delta$-isobar 
Green function
$G^0_{\Delta}(E)=(E-M^0_{\Delta}(3q))^{-1}$ transforms into the
(non-relativistic) propagator of a non-stable particle
\begin{equation}
G_R(E)=\left(E-M_{R}(E)+\frac{i}{2}\Gamma_R(E)\right)^{-1}
\label{green}
\end{equation}
and the following on-mass-shell condition should be satisfied:
\begin{equation}
\left\{ \parbox{6cm}{
\hbox{$M_R(E\!=\!M_{\Delta})=M_{\Delta}$\qquad\quad at $M_{\Delta}$=1232~MeV,}
\hbox{\ $\Gamma_R(E\!=\!M_{\Delta})=\Gamma_{\Delta}$\qquad\quad
at $\Gamma_{\Delta}$=120~MeV.}}\right.
\label{ons}
\end{equation}

The $\pi N\Delta$ vertex function Eq.(\ref{vd}) predicts the value of the
 vertex constant $f^0_{\pi N\Delta}$ and Eq.~(\ref{vd}) relates the vertex 
 form factor
 $F^0_{\pi N\Delta}(k^2)$ to the momentum distribution of quarks in the
 nucleon
\begin{equation}
f^0_{\pi N\Delta}=2\sqrt{2}f_{\pi qq}\ \ \mathrm{and}\ \ 
F^0_{\pi N\Delta}(k^2)\equiv F^0_{QM}(k^2)=e^{-k^2b^2/6}
\label{cc1}
\end{equation}
We consider the function $F^0_{QM}(k^2)$ as a quark-motivated vertex form
 factor and apply Eq.(\ref{mod})
for modification of all the meson-baryon and photon-baryon vertices.

In this simple model the decay widths $\Gamma$ of the lowest nucleon
 excitations $B=\Delta,\,R,\,N^*$ into the pion channel ($B\to\pi+N$) amount
 to 50-60\% of the experimental ones. The $\Gamma$ calculated at
 $b_{\pi}/b=0.5$ in the $^3P_0$ model are listed in Table~\ref{tab:reso}. The
 $\Delta$-isobar width $\Gamma_{\Delta}$ does not depend on the pion radius
 $b_{\pi}$, but for orbital or radial excitations ($N^*$, $R$) such a
 dependence is important. The value $b_{\pi}/b\approx0.5$ leads to an optimal
 value of the width of each resonance.

In a more realistic model the FSI in the decay channel $N+\pi$ should be taken
 into account, e.g. in terms of the unitary transformation method~\cite{lee}.
 As it was shown by Ref.~\cite{lee} the final result obtained by taking into
 account the full non-resonance interaction in the $\pi N$ channel is 
 equivalent to a small
 renormalization of the ``bare" $\pi N\Delta$ coupling constant
 Eq.(\ref{cc1}):
\begin{equation}
f^0_{\pi N\Delta}\to f_{\pi N\Delta}=Z_Rf^0_{\pi N\Delta},
\label{cc2}
\end{equation}
with $Z_R\approx1.3$. Following this prescription we obtain a very similar
 result using the quark motivated form factor~$F_{QM}$, namely $Z_R=1.37$. The
 substitution of Eq.~(\ref{cc2}) in Eqs.~(\ref{sig}) and~(\ref{vd}) does not
 violate the unitarity of final results. Thus in practice, the usage of this
 prescription is more convenient than the summation of many Born terms in the
 $\pi N$ FSI. The renormalized values of the decay widths~$\Gamma_R$ of the
 lowest baryon resonances are shown in Table~\ref{tab:reso} in comparison with
 the experimental ones $\Gamma_B$.

\begin{table}
\begin{center}
\caption{\label{tab:reso}
 Pion decay widths of baryon resonanses in the $^3P_0$ model.
}
\vspace*{3mm}
\begin{tabular}{c|c|c|c|c|c}
\hline
&&\multicolumn{4}{|c}{$B$}\\[5pt]
\cline{3-6}
&&$N(940)$&$\Delta(1232)$&$R(1440)$&$N^*(1535)$\\[5pt]
\hline
{$^3P_0$ model}&$f_{\pi NB}/f_{\pi NN}$&1&$\frac{3}{5}2\sqrt{2}$&
$\frac{3}{5}\frac{5\alpha}{\sqrt{3}}$&
$\frac{3}{5}\frac{2\sqrt{2}\beta}{\sqrt{3}}$\\[5pt]
\cline{2-6}
&$\Gamma$(MeV)&&64&47&52\\[5pt]
\hline
{Realistic model}&$\Gamma_R$(MeV)&&120&88&98\\[5pt]
\hline\hline
{exp.}&$\Gamma_B(MeV)$&&120&210-245&52-83\\[5pt]
\hline
\end{tabular}
\end{center}
\end{table}
The results in Table~\ref{tab:reso} were obtained
with the following values of parameters $\alpha$, $\beta$ and $x$ 
determined in the $^3P_0$-model:
\begin{equation}
\alpha=3a+\frac{m_{\pi}}{m_{q}}(1-\frac{a}{3}+\frac{5a^2}{9}),\quad
\beta=\frac{m_{\pi}}{m_qb}(1-\frac{2a}{3}),\quad
a=x^2(1+2x^2/3)^{-1}\ \ \mathrm{and}\ \ x=b_{\pi}/b=0.5.
\end{equation}
The widths $\Gamma$ and $\Gamma_R$ are calculated with the
standard values for the  coupling constants
$f_{\pi qq}=\frac{3}{5}f_{\pi NN}$ and $f_{\pi NN}=$ 1,
and for the scale parameter and $b=$0.6~fm.

The contribution of a virtual $\Delta$-isobar to pion-production processes is
 described by the following amplitudes (Fig.~\ref{f11}):
\begin{figure}[hp]
\begin{center}
\epsfig{file=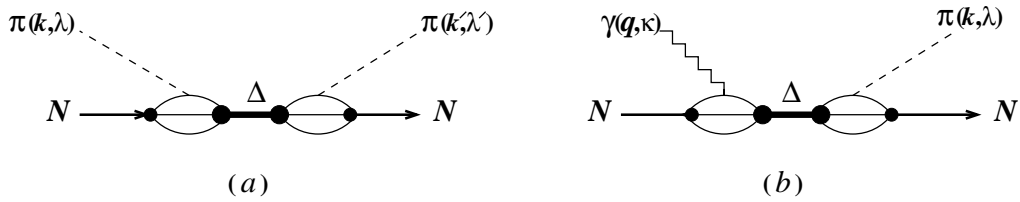,width=0.75\textwidth,clip}
\caption{\label{f11}
Resonance $\Delta$-isobar contributions to physical processes.}
\end{center}
\end{figure}

\noindent
(a) the $\pi$$N$ resonance scattering  sketched in Fig.~\ref{f11}(a) 
is written as
\begin{equation}
\frac{v_{\Delta}({\bf k^{\prime}},\lambda^{\prime})v_{\Delta}^{\dag}({\bf k},
\lambda)}{E-M_{R}(E)+\frac{i}{2}\Gamma_R(E)},
\label{pin}
\end{equation}

\noindent
(b) the pion photo-production through an intermediate $\Delta$-isobar of
 Fig.~\ref{f11}(b)
\begin{equation}
\frac{v_{\Delta}({\bf k},\lambda)w_{\Delta}^{\dag}({\bf q},
\varkappa)}{E-M_{R}(E)+\frac{i}{2}\Gamma_R(E)},
\label{intd}
\end{equation}
where the e.-m. vertex $w_{\Delta}^{\dag}({\bf q},\varkappa)$
for the M1 transition $N+\gamma\to\Delta$
\begin{equation}
w_{\Delta}^{\dag}({\bf q},\varkappa)=
3\langle N|H_{\gamma qq}^{(3)}({\bf q},\varkappa)|\Delta\rangle=
-i\left(\frac{\sqrt{2}e}{2m_q}\right)
e^{-q^2b^2/6}[{\bf S}^{\dag}\times{\bf q}]
\cdot{\b\epsilon}^{(\varkappa)}\,T_z
\label{evrt}
\end{equation}
is calculated with the elementary quark-photon coupling operator
\begin{equation}
H_{\gamma qq}^{(3)}({\bf q},\varkappa)=\frac{e^{(3)}}{2m_q}\,
e^{-i\frac{2}{3}{\bf q}\cdot{\b\rho}}\,{\b\epsilon}^{(\varkappa)}\!
\cdot\!\left(-2i{\b\nabla_{\rho}}\!+\!2{\bf q}/3-
i[\sigma^{(3)}\times{\bf q}]\right),
\label{qph}
\end{equation}
where $e^{(3)}=1/6\!+\!\tau^{(3)}_z/2$.

\subsection{Testing the model in pion photoproduction off the nucleon}
\label{ssec:pionprod}

\begin{figure}[hp]
\begin{center}
\epsfig{file=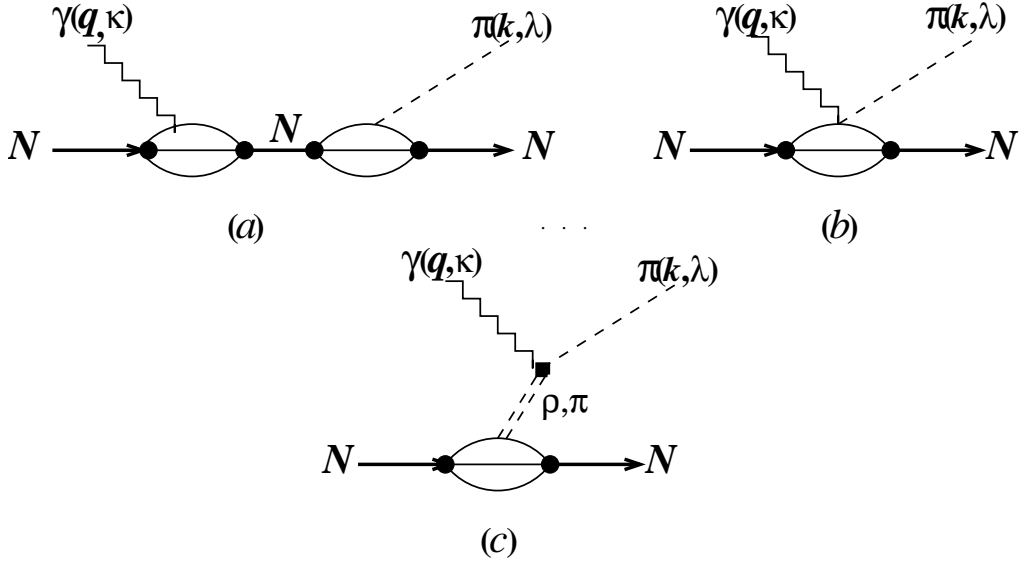,width=0.75\textwidth,clip}
\caption{\label{f12}
Non-resonance contributions to the pion photoproduction amplitude.}
\end{center}
\end{figure}

The gauge invariance and the current conservation constraint 
 has been actively used to define additional
 contributions to the photoproduction amplitude in the $\Delta$-resonance
 region. The most important ones are the nucleon-pole and pion-in-flight
 diagrams  with the seagull term added for the current conservation (see
 Fig.~\ref{f12}).

All the terms have been evaluated in the above quark approach.
The nucleon pole term depicted in Fig.~\ref{f12}(a) has the form analogous to
the pole term in Eq.~(\ref{intd})
\begin{equation}
\frac{v_{\scriptscriptstyle N}({\bf k},\lambda)w_{N}^{\dag}({\bf q},
\varkappa)}{E-M_{N}},
\label{npt}
\end{equation}
with
\begin{equation}
v_{\scriptscriptstyle N}({\bf k},\lambda)=
3\langle N(\mu^{\prime},
t^{\prime})|H^{(3)}_{\pi qq}({\bf k},\lambda)|
N(\mu_{\scriptscriptstyle P},
t_{\scriptscriptstyle P})\rangle=
i\left(\frac{5f_{\pi qq}}{3m_{\pi}}\right)
\,e^{-k^2b^2/6}\,\langle\mu^{\prime}|{\b \sigma }\cdot{\bf k}
|\mu_{\scriptscriptstyle P}\rangle\,\langle t^{\prime}|\tau_{\lambda}
|t_{\scriptscriptstyle P}\rangle,
\label{vn}
\end{equation}
and
\begin{equation}
w_{N}^{\dag}({\bf q},\varkappa)=
3\langle N(\mu_{\scriptscriptstyle P},t_{\scriptscriptstyle P})|
H_{\gamma qq}^{(3)}({\bf q},\kappa)|N(\mu,t)\rangle=
-i\left(\frac{e}{2m_q}\right)\,e^{-q^2b^2/6}
\langle\mu_{\scriptscriptstyle P}|
[{\b\sigma}^{\dag}\times{\bf q}]\cdot{\b\epsilon}^{(\varkappa)}|\mu\rangle\,
\langle t_{\scriptscriptstyle P}|\tau_z|t\rangle
\label{evn}
\end{equation}
 The contact (``seagull'') term of Fig.~\ref{f12}(b) is calculated as
matrix element
\begin{equation}
{\cal M}_{sea}({\bf q},\kappa;{\bf k},\lambda)=3\langle N|
H^{(3)}_{sea}({\bf q},\kappa;{\bf k},\lambda)|N\rangle
\label{msea}
\end{equation}
of the operator
\begin{equation}
H^{(3)}_{sea}({\bf q},\varkappa;{\bf k},\lambda)=i\lambda\, e\,
\frac{f_{\pi qq}}{m_{\pi}}\,\tau^{(3)}_{-\lambda}\,{\b\sigma}^{(3)}\!\cdot\!
{\bf\epsilon}^{(\varkappa)}\,e^{2i({\bf k}-{\bf q})\cdot{\b\rho}/3}
\label{hsea}
\end{equation}
obtained by ``gauging'' of the PV $\pi qq$ interaction Lagrangian
$ {\cal L}^{PV}_{\pi qq}=\frac{f_{\pi qq}}{m_q}
\bar\psi_q\gamma^{\mu}\gamma_5{\b\tau}\psi_q\cdot\partial_{\mu}{\b\varphi}$.
The expression for the operator is taken in Eq.~(\ref{hsea}) in a
non-relativistic limit. The final expression for the matrix element of
Eq.(\ref{msea}) can be written in the form of a series:
\begin{equation}
{\cal M}_{sea}({\bf q},\varkappa;{\bf k},\lambda)=
-ie\,\lambda\tau_{-\lambda}\,{\b\sigma}\cdot{\b\epsilon^{(\varkappa)}}
\left(\frac{5f_{\pi qq}}{3m_{\pi}}
\right)e^{-k^2b^2/6}e^{-q^2b^2/6}\nonumber\\
\times[i_0(kqb^2/3)+3\hat{\bf k}\cdot\hat{\bf q} i_1(kqb^2/3)+\dots],
\label{seag}
\end{equation}

The pion-in-flight diagram is shown in Fig.~\ref{f12}(c). Its contribution
 takes the form:
\begin{eqnarray}
{\cal M}_{\pi\pi}({\bf q},\varkappa;{\bf k},\lambda)=
ie\,\lambda\tau_{-\lambda}\left(\frac{5f_{\pi qq}}{3m_{\pi}}\right)\,
{\bf k}\cdot{\b\epsilon}^{(\varkappa)}
\frac{{\b\sigma}\cdot({\bf q}-{\bf k})}
{2(\omega_{\pi}q_0-{\bf k}\cdot{\bf q})}\,
e^{-k^2b^2/6}e^{-q^2b^2/6}\nonumber\\
\times[i_0(kqb^2/3)+3\hat{\bf k}\cdot\hat{\bf q} i_1(kqb^2/3)+\dots],
\label{pifl}
\end{eqnarray}

The $\Delta$-isobar and nucleon pole terms
 in Eqs.(\ref{intd}) and (Eq.\ref{npt})
 imply the summation over spin and isospin projections of the intermediate
 $\Delta$ or $N$ ($\mu_{\scriptscriptstyle\Delta}$,
 $t_{\scriptscriptstyle\Delta}$ or $\mu_{\scriptscriptstyle P}$,
 $t_{\scriptscriptstyle P}$ correspondingly), and thus the full expressions
 for these terms are
\begin{eqnarray}
{\cal M}_{\Delta\pi}({\bf q},\varkappa;{\bf k},\lambda)=
-\left(\frac{2\sqrt{2}f_{\pi qq}}{m_{\pi}}\right)
\left(\frac{\sqrt{2}e}{2m_q}\right)\frac{e^{-k^2b^2/6}\,e^{-q^2b^2/6}}
{\sqrt{s}-M_R(s)+\frac{i}{2}\Gamma_{\Delta}(s)}\nonumber\\
\times \sum_{\mu_{\scriptscriptstyle\Delta}}\langle\,\mu^{\prime}\,|\,
{\bf S}\cdot{\bf k}\,|\,\mu_{\scriptscriptstyle\Delta}\,\rangle\,
\langle\,\mu_{\scriptscriptstyle\Delta}\,|\,{\b\epsilon}^{(\varkappa)}
\cdot[{\bf S^{\dag}}\times{\bf q}]\,|\,\mu\,\rangle
\sum_{t_{\scriptscriptstyle\Delta}}\langle\,t^{\prime}\,|\,T_{-\lambda}\,|\,
t_{\scriptscriptstyle\Delta}\,\rangle\,
\langle\,t_{\scriptscriptstyle\Delta}\,|\,T^{\dag}_z|\,t\,\rangle
\label{dpol}
\end{eqnarray}
and
\begin{eqnarray}
{\cal M}_{N\pi}({\bf q},\varkappa;{\bf k},\lambda)=
-\left(\frac{5f_{\pi qq}}{3m_{\pi}}\right)
\left[\frac{1}{6}+(-1)^{t-1/2}\frac{5}{6}\right]\,\left(\frac{e}{2m_q}\right)
\frac{e^{-k^2b^2/6}\,e^{-q^2b^2/6}}{\sqrt{s}-M_N}\nonumber\\
\times \sum_{\mu_{\scriptscriptstyle P}}\langle\,\mu^{\prime}\,
|\,{\b\sigma}\cdot{\bf k}\,|\,\mu_{\scriptscriptstyle P}\,\rangle\,
\langle\,\mu_{\scriptscriptstyle P}\,|\,{\b\epsilon}^{(\varkappa)}
\cdot[{\b\sigma}\times{\bf q}]\,|\,\mu\,\rangle
\sum_{t_{\scriptscriptstyle P}}\langle\,t^{\prime}\,|\,
\tau_{-\lambda}\,|\,t_{\scriptscriptstyle P}\,\rangle\,
\langle\,t_{\scriptscriptstyle P}\,|\,\tau_z|\,t\,\rangle
\label{npol}
\end{eqnarray}

\begin{figure}[hpbt]
\begin{center}
\epsfig{file=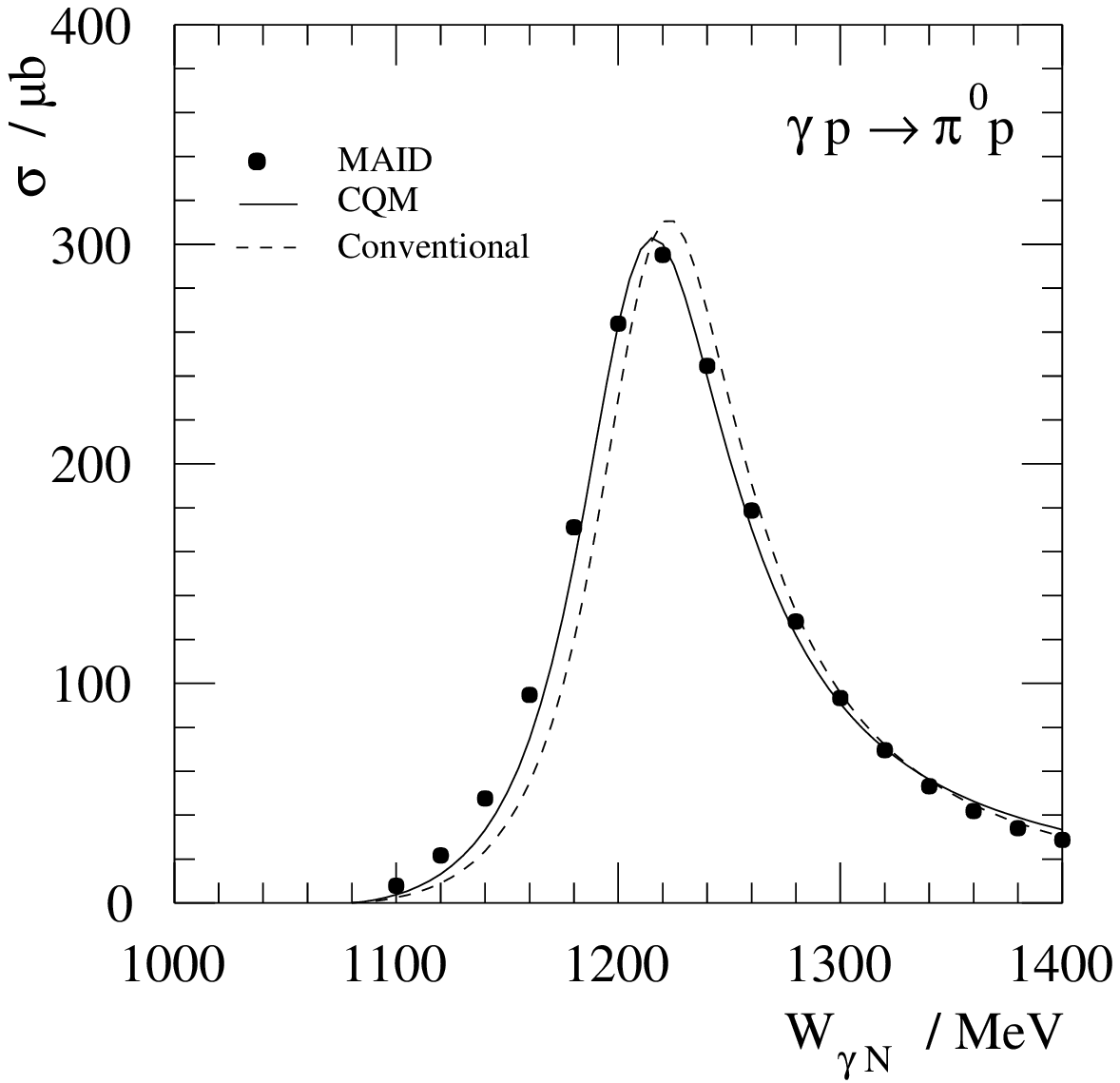,width=0.45\textwidth,clip}\hspace*{10mm}
\epsfig{file=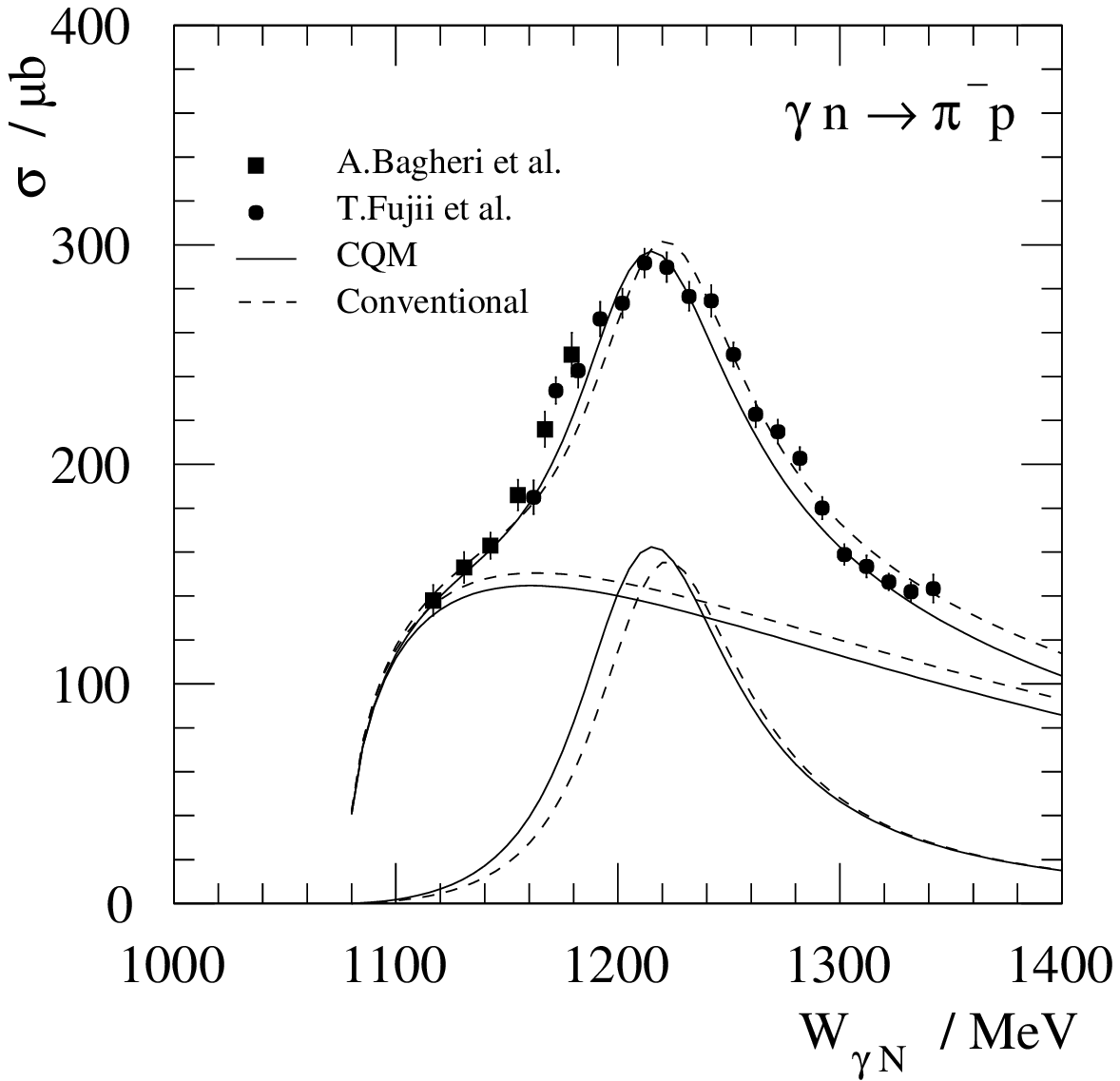,width=0.45\textwidth,clip}
\caption{\label{f13}
Total cross section for $\pi^\mathrm{o}$ photoproduction off the
proton (left) and for $\pi^-$ photoproduction off the neutron (right).
Calculations in conventional (dash line) and quark (solid line)
models. The data are taken from Refs.~\cite{mai} for $\pi^\mathrm{o}$
and~\cite{bag} for $\pi^-$, respectively. 
}
\end{center}
\end{figure}

The numerical results on the energy dependence of the $\pi^\mathrm{o}$ and
 $\pi^-$ photoproduction cross sections are shown in Fig.~\ref{f13} in
comparison with the conventional model results in which the phenomenological
 $\pi N$-coupling constants and form factors are used for the same hadron
 vertices as shown in Figs.~\ref{f11} and~\ref{f12}.  One can see a good
 agreement between results of both models and the data.  For the charged pion
 $\pi^-$ the contributions of two main terms, the $\Delta$-resonance and
 seagull, are shown separately in Fig.~\ref{f13} on the right panel.

\begin{figure}[htbp]
\begin{center}
\epsfig{file=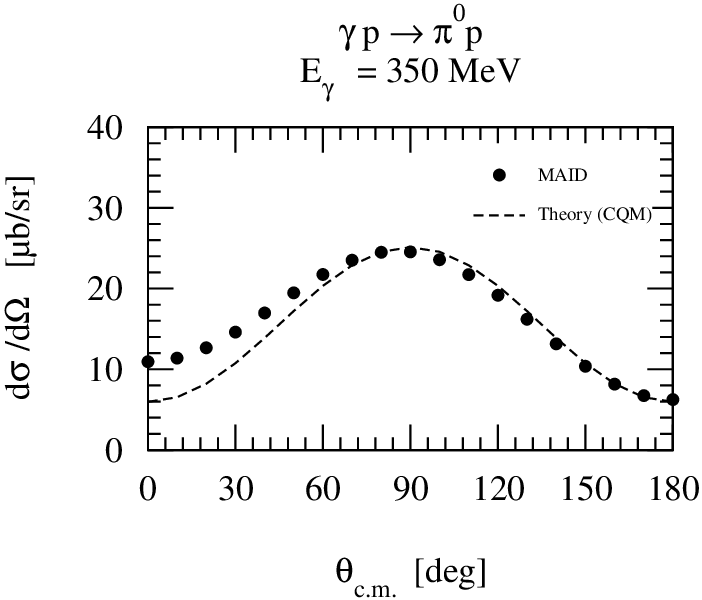,width=0.45\textwidth,clip}\hspace*{10mm}
\epsfig{file=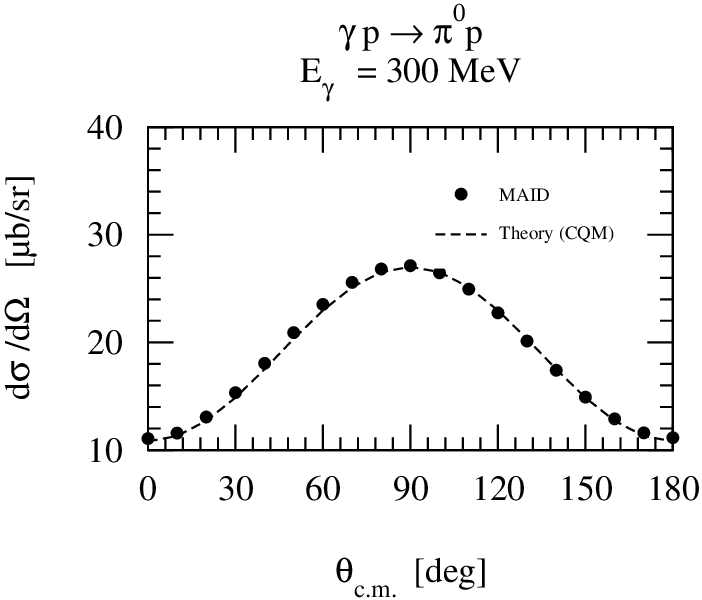,width=0.45\textwidth,clip}
\caption{\label{f15}
Differential cross section for $\pi^\mathrm{o}$
photoproduction in comparison to the data from Refs.~\cite{bag,mai}}
\end{center}
\end{figure}

\begin{figure}[hbtp]
\begin{center}
\epsfig{file=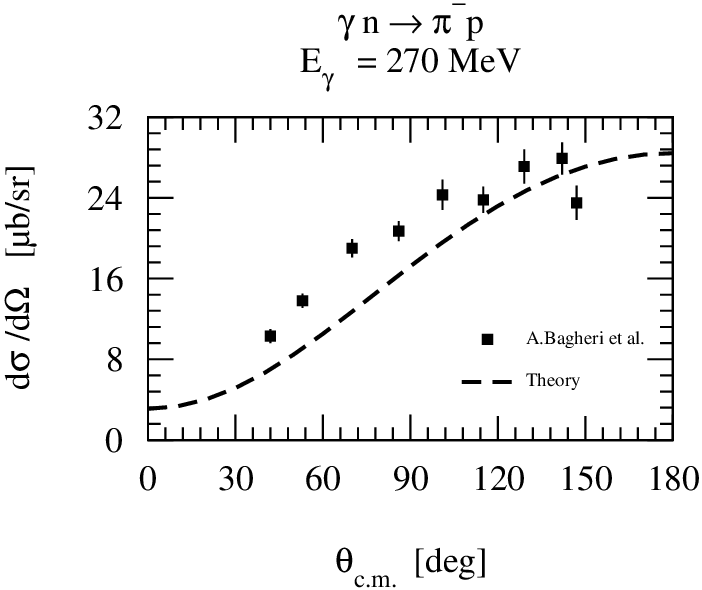,width=0.45\textwidth,clip}\hspace*{10mm}
\epsfig{file=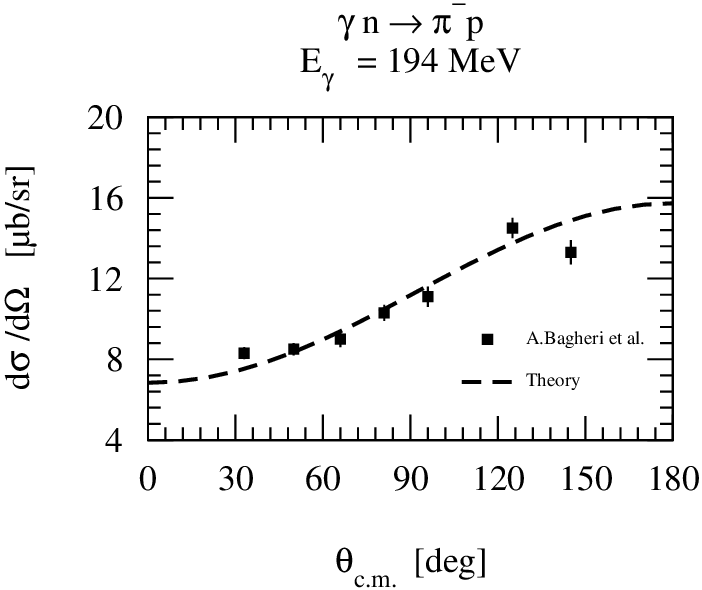,width=0.45\textwidth,clip}
\caption{\label{f18}
Differential cross section for $\pi^-$
photoproduction in comparison to the  data from Ref.~\cite{ben}. }
\end{center}
\end{figure}

For the reaction $\gamma p\to\pi^{0}p$ the differential cross section are
 shown in Fig.~\ref{f15} for photon energies of $E_{\gamma}$=300 and 350~MeV,
 while the for the $\gamma n\to\pi^{-}p$ reaction energies of $E_{\gamma}$=194
 and 270~MeV have been selected (Fig.~\ref{f18}).

The CQM is seen to lead to reasonable results for pion photoproduction off the
 nucleon, and thus this model can be employed to describe similar processes
 off the deuteron, where one can expect this model to predict new effects
 related to the $N\Delta$ FSI at short ranges.

\section{Pion photoproduction off the deuteron}
\label{sec:deuteron}

The quark model calculations of cross sections for pion photoproduction off
 the deuteron in the $\Delta$ resonance region have been performed in the
 framework of the Moscow-T\"ubingen model for $NN$ interaction.  The results
 are similar to the conventional model results without $N\Delta$ FSI. But
 taking into consideration the short-range $N\Delta$ FSI leads to an
 additional effect of suppression of the forward-angle $\pi^\mathrm{o}$
 photoproduction similar to the suppression obtained in Ref.~\cite{lev2} due
 to the $NN$ FSI.

\begin{figure}[hbtp]
\begin{center}
\epsfig{file=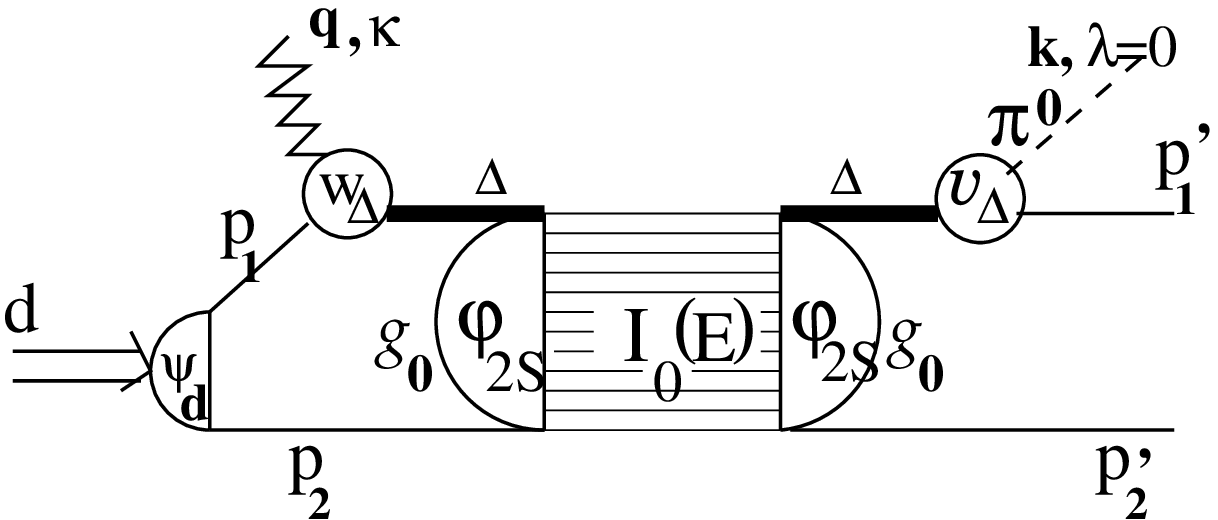,width=0.42\textwidth,clip}\hspace*{10mm}
\epsfig{file=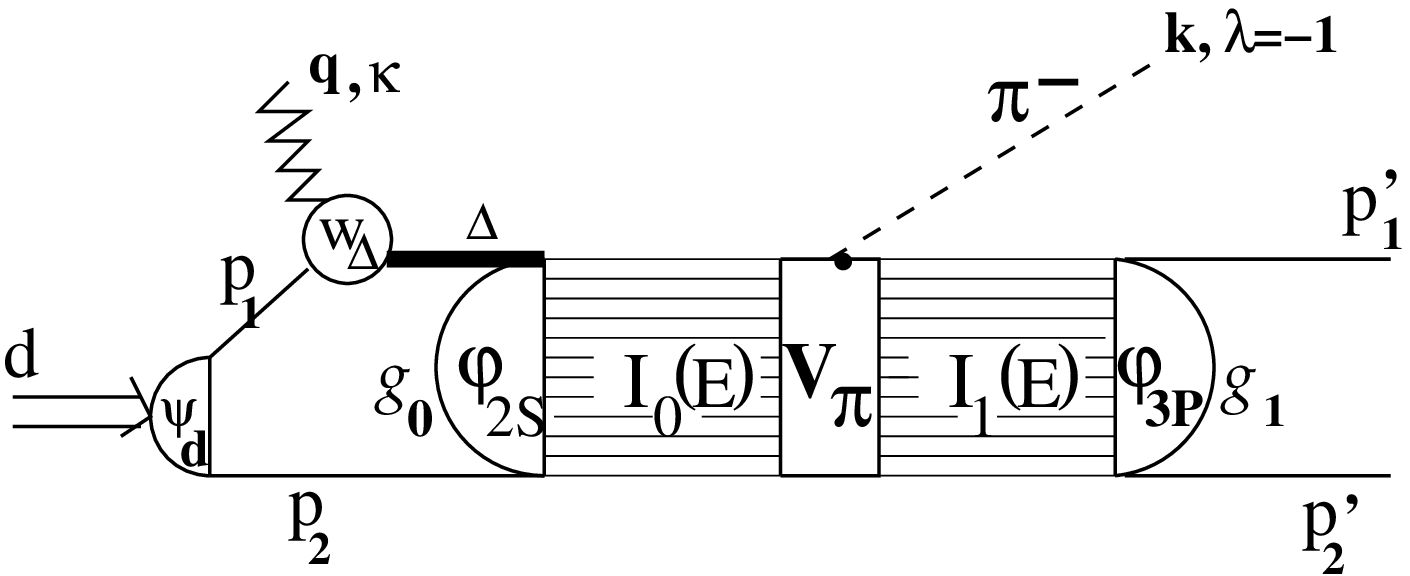,width=0.42\textwidth,clip}
\caption{\label{f22}
 The dressed-bag model. The main blocks of the
loop diagrams of $N\Delta$ FSI for $\pi^\mathrm{o}$ (left) and $\pi^-$ (right)
photoproduction.}
\end{center}
\end{figure}
The expressions for the loop diagrams of Figs.~\ref{f8}a) and~\ref{f9} are
 illustrated by the expanded diagrams in Fig.~\ref{f22} for better
 visualization of the main blocks of the expressions.

Our calculation of the $\pi^\mathrm{o}$ photoproduction (Fig.~\ref{f22}, left)
 use a loop integral which can be expressed in the form:
\begin{eqnarray}
{\cal M}_{d\gamma\to\pi^0np}({\bf q},\varkappa;{\bf k},\lambda;
{\bf p}^{\prime})=\int \frac{d^3p}{(2\pi)^3}\Psi_d({\bf p})
w_{\scriptscriptstyle\Delta}({\bf q},\varkappa)
\frac{g_0\,\varphi_{\scriptscriptstyle 2S}({\bf p}\!+\!{\bf q}/2)}
{W_{\scriptscriptstyle\Delta}({\bf p})-
M_R(W_{\scriptscriptstyle\Delta}({\bf p}))+
\frac{i}{2}\Gamma_R(W_{\scriptscriptstyle\Delta}({\bf p}))}
\nonumber\\
\times I_0(s^{1/2})\,\frac{g_0\,
\varphi_{\scriptscriptstyle 2S}({\bf p^{\prime}}\!+\!{\bf k}/2)}
{W^{\prime}_{\scriptscriptstyle\Delta}({\bf p}^{\prime})-
M_R(W^{\prime}_{\scriptscriptstyle\Delta}({\bf p}^{\prime}))+
\frac{i}{2}
\Gamma_R(W^{\prime}_{\scriptscriptstyle\Delta}({\bf p}^{\prime}))}
v_{\scriptscriptstyle\Delta}({\bf k},\lambda),
\label{gpi0}
\end{eqnarray}
for $\lambda{=}$0, where $s^{1/2}=\sqrt{M_d(M_d+2E_{\gamma})} 
\approx M_d+E_{\gamma}$ is the mass
 of the initial $\gamma+d$ state. The energies of the off-shell $\Delta$ in
 the loop ($W_{\Delta}$) and in the final $N\Delta$ state
 ($W_{\Delta}^{\prime}$) are given by the expressions
\begin{eqnarray}
W_{\scriptscriptstyle\Delta}({\bf p})\equiv
s^{1/2}_{\scriptscriptstyle\Delta}({\bf p}),\quad
s_{\scriptscriptstyle\Delta}({\bf p})=
(M_d-E_p+E_{\gamma})^2-({\bf p}+{\bf q})^2,\nonumber\\
W^{\prime}_{\scriptscriptstyle\Delta}({\bf p}^{\prime})\equiv
s^{1/2}_{\scriptscriptstyle\Delta}({\bf p}^{\prime}),\quad
s_{\scriptscriptstyle\Delta}({\bf p}^{\prime})=
(M_{\scriptscriptstyle N}+\omega_{\pi})^2+
2\omega_{\pi}(E^{\prime}_p-M_{\scriptscriptstyle N}),\nonumber\\
E_p=\sqrt{M^2_{\scriptscriptstyle N}+{\bf p}^2}\ \ \mathrm{and}\ 
E^{\prime}_p=\sqrt{M^2_{\scriptscriptstyle N}+
({\bf p}^{\prime}+({\bf q}-{\bf k})/2)^2}.
\label{sd}
\end{eqnarray}
They depend on relative momenta of nucleons in the deuteron ${\bf p}=({\bf
 p}_1-{\bf p}_2)/2$ and in the final state ${\bf p}^{\prime}=({\bf
 p}^{\prime}_1-{\bf p}^{\prime}_2)/2$ respectively (see the momentum
 designations in Fig.~\ref{f22}).  Note that the above expression for the
 energy $E_p$ of the second nucleon (with momentum ${\bf p}_2=-{\bf p}$) 
 implies
 that the nucleon-spectator is proposed to be on its mass shell.

For the $\pi^-$ photoproduction the loop integral for the right
 diagram in Fig.~\ref{f22} is expressed in a similar form
\begin{eqnarray}
{\cal M}_{d\gamma\to\pi^-np}({\bf q},\varkappa;{\bf k},\lambda;
{\bf p}^{\prime})=\int \frac{d^3p}{(2\pi)^3}\Psi_d({\bf p})
w_{\scriptscriptstyle\Delta}({\bf q},\varkappa)
\frac{g_0\,\varphi_{\scriptscriptstyle 2S}({\bf p}\!+\!{\bf q}/2)}
{W_{\scriptscriptstyle\Delta}({\bf p})-
M_R(W_{\scriptscriptstyle\Delta}({\bf p}))+
\frac{i}{2}\Gamma_R(W_{\scriptscriptstyle\Delta}({\bf p}))}
\nonumber\\
\times I_0(s^{1/2})\,V_{\pi}(k,\lambda)\,I_1(s^{1/2})\,g_1\,
\varphi_{\scriptscriptstyle 3P}({\bf p}^{\prime}),
\ \ \mathrm{for}\  \lambda{=}-1,
\label{gpim}
\end{eqnarray}
with an additional dressed-bag transition vertex $V_{\pi}(k,\lambda)$ 
instead of the $\pi N\Delta$ vertex 
$v_{\scriptscriptstyle\Delta}({\bf k},\lambda)$, where the matrix element
\begin{equation}
V_{\pi}(k,\lambda{=}\!-\!1)=-\left(\frac{44\sqrt{2}f_{\pi qq}}
{45m_{\pi}b}\right)
\left(\frac{\omega_{\pi}}{2m_q}+\frac{k^2b^2}{6}\right)\,e^{k^2b^2/6}
\label{vpim}
\end{equation}
  for  the transition $D_0(s^6)\to d_1(s^5p)+\pi(l_{\pi}{=}0)$ has
  been as usually taken from the $^3P_0$ model in the limit~(Eq.\ref{lim}).

 In Eqs.~(\ref{gpi0}) and~(\ref{gpim}) the vertex constants $g_0$ and $g_1$ are
 the algebraic values from Eq.~(\ref{g}) from $S$- and $P$-wave $N\Delta$ and
 $NN$ channels, respectively. Similarly, the pole factors $I_0(s^{1/2})$ and
 $I_1(s^{1/2})$ are functions (\ref{pade}) with parameters given by
 Eq.(\ref{param}). The $S$-wave deuteron wave function $\Psi_d({\bf p})$ in
 the loop integrals [Eqs.(\ref{gpi0}) and (\ref{gpim})] has been taken from
 the dressed-bag model~\cite{kuk} directly.  The $\pi N\Delta$ and $\gamma
 N\Delta$ vertices $v_{\scriptscriptstyle\Delta}({\bf k},\lambda)$ and
 $w_{\scriptscriptstyle\Delta}({\bf q},\varkappa)$ in the integrand of
 Eqs.~(\ref{gpi0}) and~(\ref{gpim}) are defined in Eqs.~(\ref{vd}) and
 (Eq.\ref{evrt}) (see also renormalization conditions in (Eq.\ref{mod}) and
 (Eq.\ref{cc2})).  The summation over spin and isospin projections $\mu$,
 $\mu^{\prime}$, $t$ and $t^{\prime}$ of nucleons and $\Delta$s with
 Clebsch-Gordan coefficients in Eqs.~(\ref{gpi0}) and~(\ref{gpim}) is omitted
 here to simplify reading.

\begin{figure}[hbtp]
\begin{center}
\epsfig{file=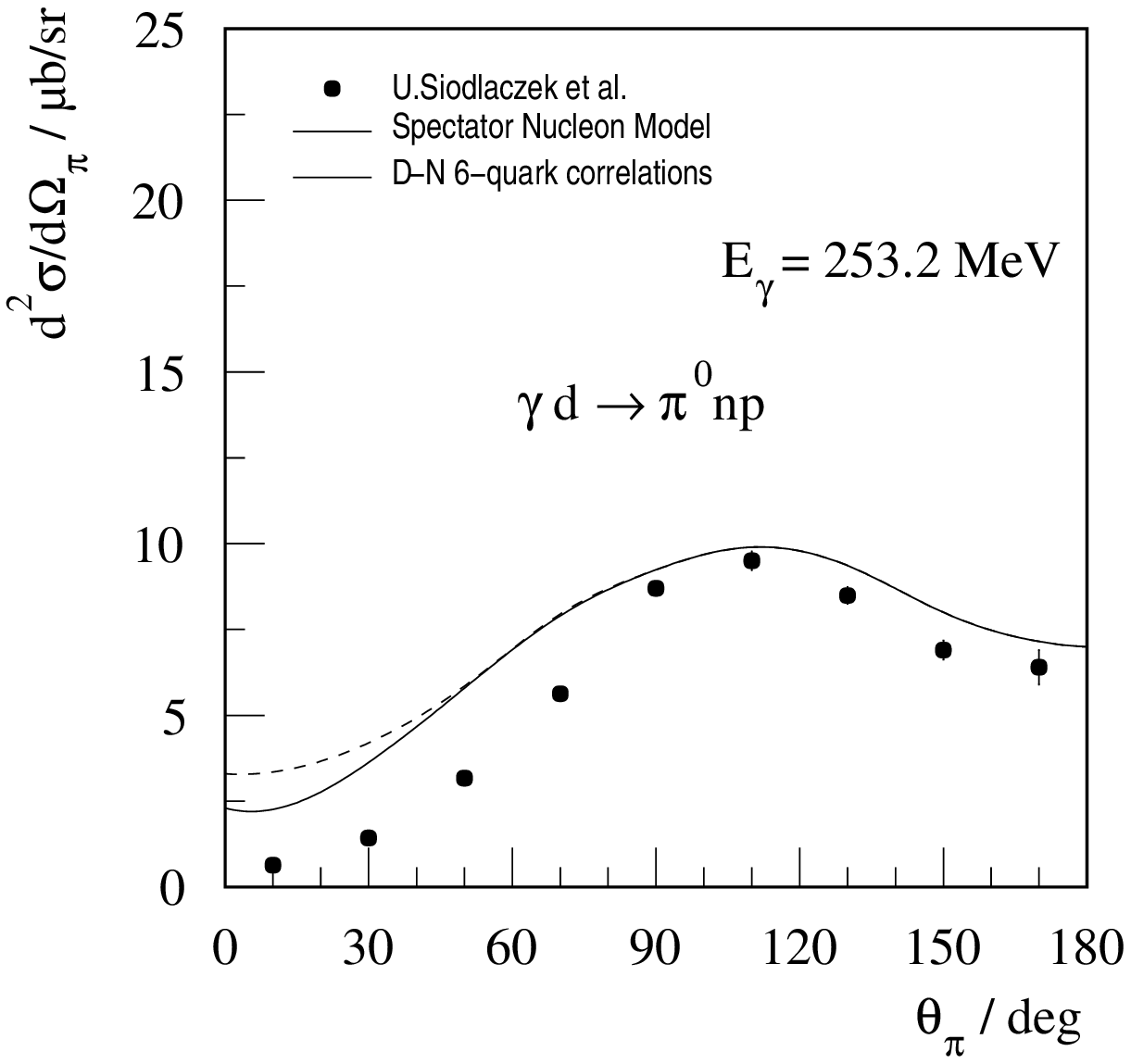,width=0.28\textwidth,clip}\hspace*{7mm}
\epsfig{file=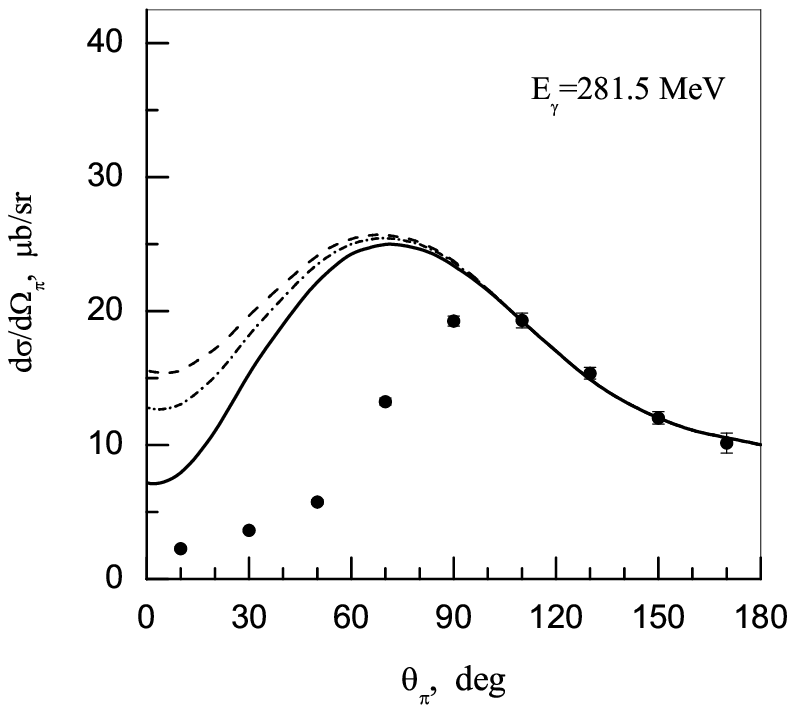,width=0.33\textwidth,clip}\hspace*{7mm}
\epsfig{file=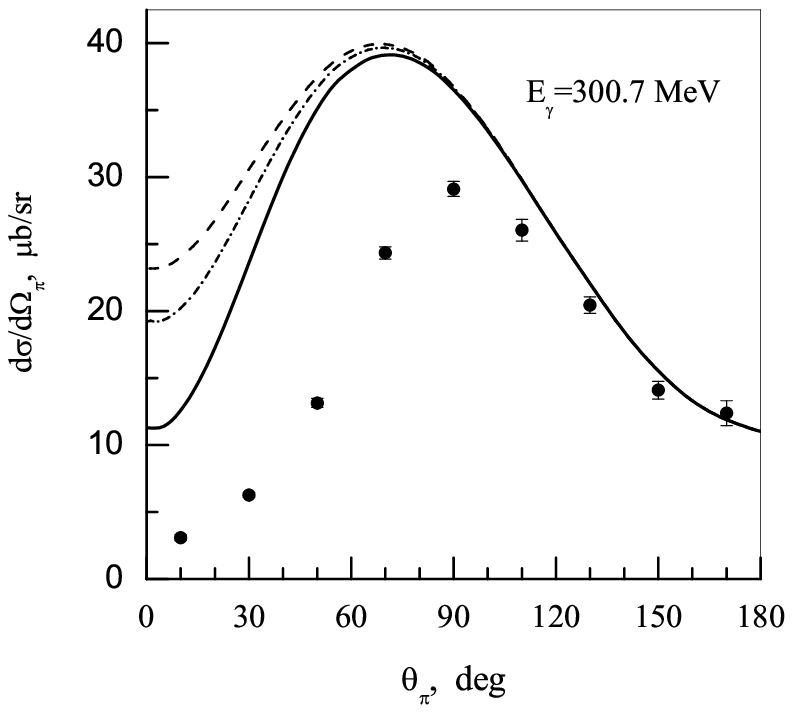,width=0.33\textwidth,clip}
\caption{\label{f19}
  The differential cross sections for
  $\pi^\mathrm{o}$ photoproduction off the deuteron~\cite{sio} 
  taken at 3 photon energies ($E_{\gamma}$ is in the c.m. of $\gamma-d$
  system) compared to the quark model predictions. 
  Results for the quasi-free mechanism of photoproduction are shown by 
  the dashed line, while the combined effect of the quasi-free mechanism 
  and the Born diagram of the $N\Delta$ FSI is shown by the dash-dot and 
  solid lines (for the renormalized value of the coupling constant 
  $g_{\scriptscriptstyle \Delta\sigma q}$ 
  from Eq.~(\ref{g}) and for the non-renormalized value
  $g_{\scriptscriptstyle \Delta\sigma q}=g_{\scriptscriptstyle N\sigma q}$ 
  respectively).
}
\end{center}
\end{figure}

\begin{figure}[hpb]
\begin{center}
\epsfig{file=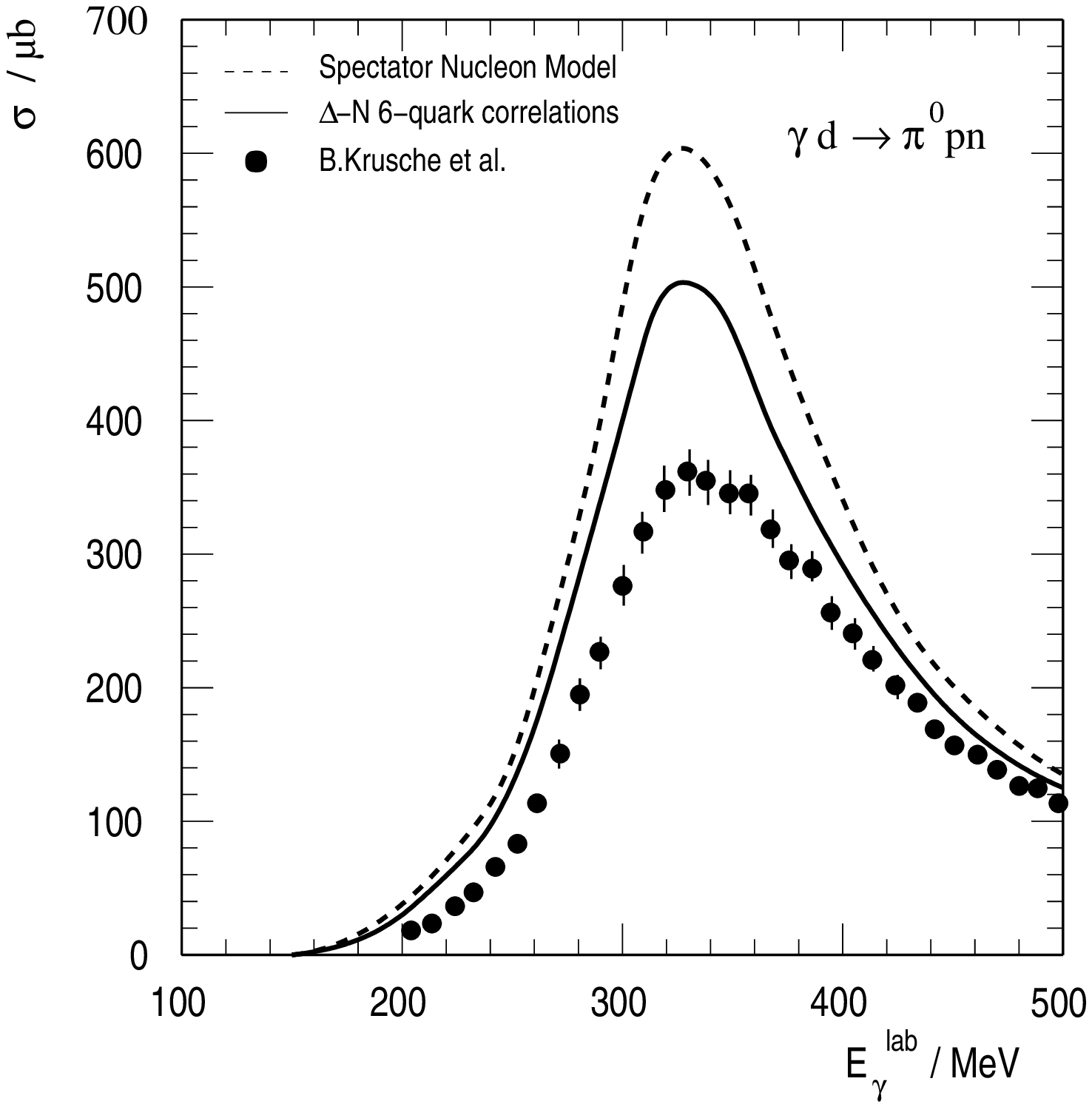,width=0.4\textwidth,clip}\hspace*{10mm}
\epsfig{file=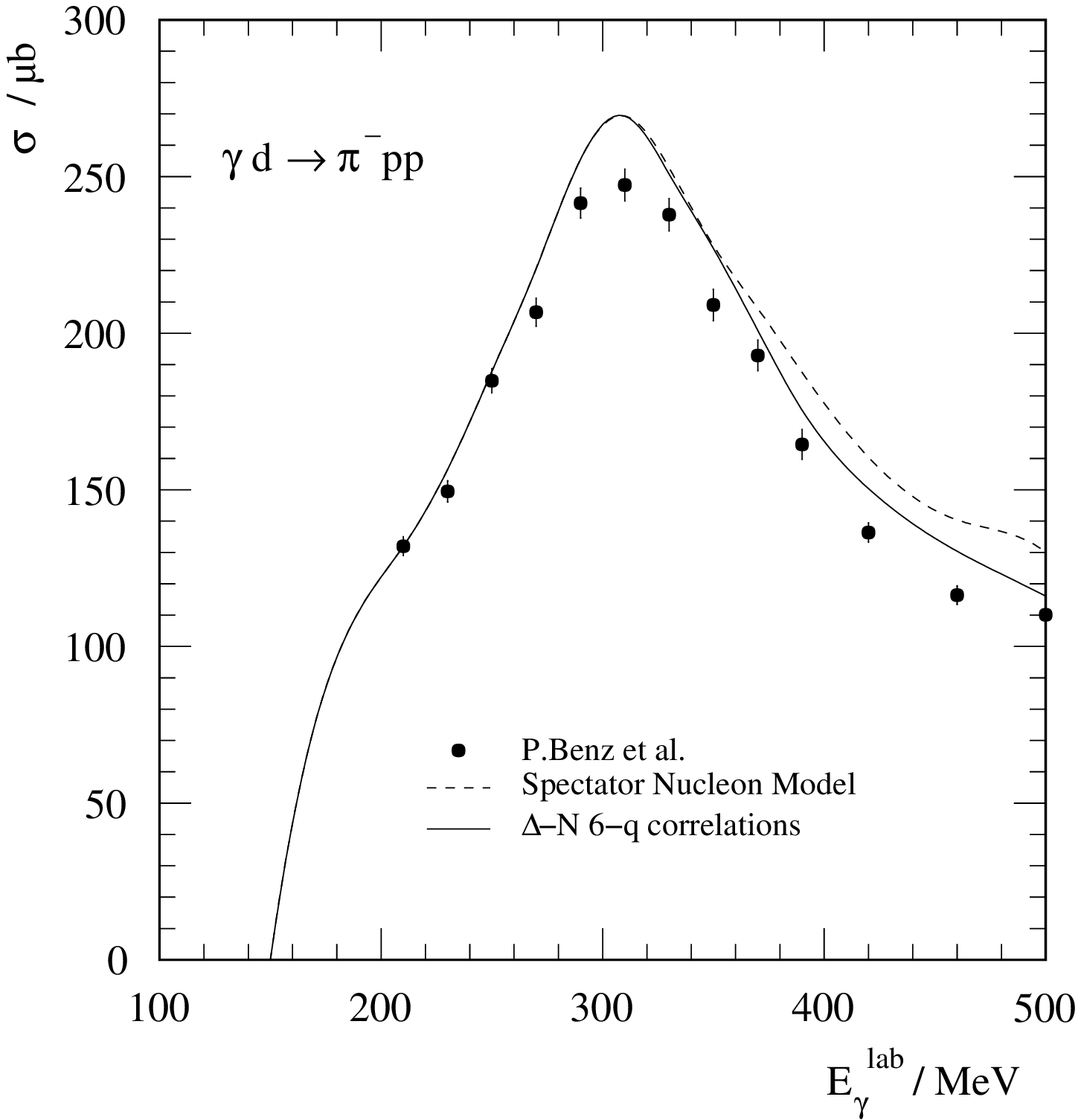,width=0.4\textwidth,clip}
\caption{\label{f20}
Left:  The total cross sections for
  $\pi^\mathrm{o}$ photoproduction off the deuteron~\cite{kru} in comparison
to those calculated in the quark
  model (full lines). 
Right:   Total cross section of $\pi^-$ photoproduction.
  Data are from Ref.~\cite{ben}.}
\end{center}
\end{figure}
The results of the calculations with and without $N\Delta$ FSI for the $\gamma
 d\to\pi^\mathrm{o}np$ are shown in Fig.~\ref{f19} for the differential cross
 sections taken at three photon energies. Results for two variants of FSI are
 shown: the dash-dot curve corresponds to the value of effective coupling 
 constant $g_{\scriptscriptstyle \Delta\sigma q}$ obtained by renormalization
 of the $NN$ parameters (\ref{param}) with the algebraic factors 
 $C^{\scriptscriptstyle \Delta}_{\scriptscriptstyle ST}/
 C^{\scriptscriptstyle N}_{\scriptscriptstyle ST}$ 
 of Eq.~(\ref{g}), while the solid curve corresponds to the non-renormalized
 values of the parameters (\ref{param}). From these two curves it is possible 
 to make rough estimates of model dependece of the results.   

The total cross sections for the
 $\pi^\mathrm{o}np$ and the $\pi^-pp$ channels are compared in
 Fig.~\ref{f20}. One can see that including the $N\Delta$ FSI in terms of the
 six-quark bag diagram of Fig.~\ref{f8}(a) leads to a detectable effect for
 both the differential and integral cross sections of $\pi^\mathrm{o}$
 production. It is also seen that the contribution of the $N\Delta$ FSI to the
 $\pi^0$ photoproductin has a resonance character, and out of the resonance
 region (e.g. at $E_{\gamma}=$253.2 Mev in the c.m. of $\gamma-d$ system)
 the contribution becomes a negligibly small quantity. 

Still, a discrepancy to the data remains.  However, for the
 $\pi^-$ photoproduction (Fig.~\ref{f20}) the FSI is not so important as for
 the $\pi^\mathrm{o}$. These results are consistent with our qualitative
 consideration in Sect.\ref{ssec:srinter}.

\section{Conclusions}
\label{sec:conclusion}

The dressed-bag model predictions for the contribution of $N\Delta$ FSI to the
 $d(\gamma,\pi^0)np$ and $d(\gamma,\pi^-)pp$ cross sections are qualitatively
 consistent with the data, but full calculations with the off-shell amplitudes
 of $NN$ and $N\Delta$ FSI are necessary to obtain a quantitative description
 of the cross sections.

It is well known~\cite{lev2} that the main correction to the quasi-elastic
 (spectator) mechanism of $\pi^\mathrm{o}$ photoproduction off the deuteron is
 due to $n\!-\!p$ rescattering in the final state (diagrams in
 Fig.~\ref{f1}b).  The present calculation demonstrates that the cross section
 of the reaction $\gamma d\to \pi^\mathrm{o}np$ is also sensitive to the
 final-state $N\Delta$ interaction at least in the $\Delta$-resonance region
 for forward pion angles.

Our results can only be considered as a first step towards a quantitative
 description of photoproduction processes with taking into account both quark
 and meson degrees of freedom. Now we can only state some qualitative
 implications of the study:
\begin{itemize}
  \item[i)] It is shown that the CQM (supplemented by a simple dressing
  procedure based on the $^3P_0$ model) not only leads to a
  workable approach to the pion photoproduction off the nucleon, but it can
  also be used for evaluations of photoproduction cross sections on lightest
  nuclei, e.g. D and $^3$He.
\item[ii)] The dressed-bag model of short-range $NN$ and $N\Delta$
  interaction based on an analysis of the configuration mixing in the overlap
  region predicts a destructive interference of the $N\Delta$ FSI loop diagram
  (Fig.~\ref{f8}a) with the quasi-elastic photoproduction Born amplitudes
  calculated within the CQM.
\item[iii)] This model predicts new types of $NN$ and $N\Delta$ short-range
  correlations that would be desirable to investigate in photoproduction
  reactions on lightest nuclei, e.g. in polarization experiments as it was
  proposed in Sect.~\ref{ssec:srinter} for the mechanism shown in
  Fig.~\ref{f9} (another polarization experiments was recently proposed in
  Ref.~\cite{rek}).
\end{itemize}

The most important conclusion which can be derived from the above CQM
 consideration is that the short-range baryon-baryon interaction is determined
 by a bag-like six-quark state. Despite the fact that the bag  has its own 
 meson cloud this short-range interaction has a non-Yukawa form and directly 
 leads to the $s$-channel singularity of amplitudes. 
In experiments at more high energies
 $E_{\gamma}\gtrsim$1-2 GeV, where the $t$-channel singularities become
 extinct because of form factors, the $s$-channel singularities should lead
 to more pronounced contributions to observable effects.

\section{Acknowledgment}

We gratefully acknowledge very useful discussions with Profs. V. Neudatchin,
M. Rosina, B. Golli and Fl. Stancu and with Drs. V. Lyubovitskij, 
A. Machavariani and V.N. Pomerantsev.  This work
 was supported by funds from the Russian Foundation for Basic Research (grants
 01-02-04015 and 02-02-16612) and the Deutsche Forschungsgemeinschaft
 (Fa67/20-1, GRK683, Gr1084/3, and He2171/1).

\end{document}